\documentclass{agujournal2018}
\usepackage{apacite}
\usepackage{url} 

\draftfalse

\journalname{arXiv}

\begin{document}

%
%


\title{The Effects of Solar Wind Dynamic Pressure on the Structure of the Topside Ionosphere of Mars  
}

%
%




\authors{Z. Girazian\affil{1}, 
		 J. Halekas\affil{1},
		 D. D. Morgan\affil{1},
         A. J. Kopf\affil{1},
         D. A. Gurnett\affil{1},
         F. Chu\affil{1}}

\affiliation{1}{Department of Physics and Astronomy, University of Iowa, Iowa City, Iowa}




\correspondingauthor{Zachary Girazian}{zachary-girazian@uiowa.edu}




\begin{keypoints}
\item The topside ionosphere is globally depleted of plasma during times of high solar wind dynamic pressure.
\item The topside ionosphere responds to changes in solar wind dynamic pressure in regions of both weak, and strong, crustal magnetic fields.
\item Day-to-night plasma transport in the high-altitude nightside ionosphere is reduced during times of high solar wind dynamic pressure.

\end{keypoints}

%
%


\begin{abstract}

We use Mars Atmosphere and Volatile EvolutioN observations of the upstream solar wind, and Mars Express observations of ionospheric electron densities and magnetic fields, to study how the topside ionosphere ($>$ 320 km) of Mars is affected by variations in solar wind dynamic pressure. We find that high solar wind dynamic pressures result in the topside ionosphere being depleted of plasma at all solar zenith angles, coincident with increased induced magnetic field strengths. The depletion of topside plasma in response to high solar wind dynamic pressures is observed in both weak and strong crustal magnetic field regions. Taken together, our results suggest that high solar wind dynamic pressures lead to ionospheric compression, increased ion escape, and reduced day-to-night plasma transport in the high-altitude nightside ionosphere.

\end{abstract}




%
%

\section{Introduction}
\label{intro}

	At Mars, the solar wind interacts directly with the extended atmosphere and ionosphere to form a magnetic barrier that deflects the solar wind around planet \citep{nagy2004,ma2004,halekas2017a}. Through this interaction, the solar wind drives heating, acceleration, and escape of ionospheric plasma \citep{dubinin2011,dong2015,brain2016,halekas2016,cravens2017,dong2017,dubinin2017b,collinson2018,fowler2018}. Thus, characterizing the structure of the topside ionosphere, and its variations with upstream solar wind conditions, is important for understanding the ionosphere-solar wind interaction and the physical processes that control ion escape at Mars. The topside ionosphere ($>\sim$200 km) has been a topic of interest ever since the Viking Lander observations in 1978, which models could reproduce only by invoking substantial plasma outflows \citep{hanson1977,chen1978,kar1996a,fox1997,fox2009a}. 
	
	Observational studies detailing how the topside ionosphere is affected by upstream solar wind conditions have, to date, primarily focused on case studies during impulsive space weather events \citep{crider2005,dubinin2009,perez2009,edberg2009b,edberg2010,morgan2014,withers2012,opgen2013,jakosky2015b,withers2016,duru2017b,harada2017,luhmann2017,ma2017,ramstad2017,sanchez2017,thampi2018}. These studies have found that space weather events - such as interplanetary coronal mass ejections, co-rotating interaction regions, and solar wind pressure pulses - compress the upper ionospheric boundary and increase the induced magnetic field strength in the ionosphere. They have also found that, during the events, plasma densities in the topside ionosphere are highly-variable and depleted compared to quiescent conditions, which is often interpreted as being the result of ionospheric compression or increased ion outflow \citep{dubinin2009,fox2009,edberg2009b,edberg2010,opgen2013,mendillo2017,duru2017b,ma2017,wu2019}.

	Many of these studies were limited to using only approximations of upstream solar wind parameters because the first dedicated solar wind monitors arrived at Mars only in 2014 as part of the Mars Atmosphere and Volatile EvolutioN (MAVEN) mission \citep{jakosky2015}. These studies were also focused on individual space weather events so they were unable to perform statistical analyses concerning the global ionospheric response to upstream solar wind conditions. Recently, \citet{dubinin2018} used MAVEN data to complete the most robust statistical analysis to date, which focused on O$^+$ densities in the topside ionosphere. They found that O$^+$ densities above $\sim$450 km decreased with increasing solar wind dynamic pressure.
	
	The crustal magnetic fields at Mars are also known to significantly affect the solar wind interaction and the structure of the ionosphere \citep{nagy2004,brain2005,andrews2015,nemec2016,flynn2017,nemec2019,withers2019}. Although recent work suggests that magnetic topology plays a significant a role \citep{fowler2019}, it is still an open question as to how the crustal fields affect the response of the topside ionosphere to variations in upstream solar wind conditions.
	
	Motivated by these results, and the lack of a complete understanding of how the ionosphere responds globally to upstream solar wind conditions, we utilize MAVEN observations of the upstream solar wind dynamic pressure, and Mars Express (MEX) observations of ionospheric electron densities, to show how the solar wind dynamic pressure affects the global distribution of plasma in the topside ionosphere of Mars. We also investigate how crustal magnetic fields influence the response of the topside ionosphere to variations in the solar wind dynamic pressure.

  \section{Data}

  \subsection{Electron Densities and Magnetic Fields from Mars Express}

	We use electron densities obtained from electron plasma oscillations observed during radar soundings made by the Mars Advanced Radar for Subsurface and Ionosphere
    Sounding (MARSIS) instrument on MEX \citep{picardi2004,gurnett2005,gurnett2008}. When MARSIS sounds the ionosphere, it transmits pulses at 160 different frequencies between 0.1 and 5.4 MHz and records the return echoes that are reflected off the ionosphere. The frequency sweeps are repeated every 7.54 seconds for $\sim$40 minutes when MEX is below $\sim$1500 km during its seven hour orbit. 
	Each frequency sweep produces an ionogram - the echo intensity as a function of frequency and time delay. 
	
	In this study, we use electron densities extracted from ionograms by measuring the frequency spacing between vertical lines that appear between 0.1 and 2.0 MHz. These vertical lines are echoes caused by the excitation of electron plasma oscillations during ionospheric sounding \citep{gurnett2005,gurnett2008}. The electron densities are measured locally at spacecraft location and detailed discussions that describe how they are derived are presented in \citet{duru2008} and \citet{gurnett2005,gurnett2008}.  
	
	The local electron densities can only be detected when the plasma flow velocity is less than 160 km s$^{-1}$, the plasma temperature is below 80,000 K, and the electron density exceeds $\sim$10 cm$^{-3}$ \citep{duru2008,duru2009}. Given these constraints, the local electron density measurements are made within the ionosphere and below the magnetosheath where plasma flow velocities typically exceed the 160 km s$^{-1}$ threshold \citep{duru2009}.
	
    
    We also use MARSIS measurements of the local magnetic field strength. The magnetic field is extracted from ionograms by measuring the spacing between horizontal lines that sometimes appear at harmonics of the local electron cyclotron frequency. They are caused by the excitation of electron plasma gyrations around local magnetic field lines during ionospheric soundings. Details of how the magnetic fields are derived are presented in \citet{gurnett2008} and \citet{akalin2010}.
    
  \subsection{Upstream Solar Wind Conditions from MAVEN}

	We use measurements of the solar wind velocity ($V_{sw}$) and mass density ($\rho_{sw}$) from the MAVEN Solar Wind Ion Analyzer (SWIA) \citep{halekas2015a,halekas2017}. The data are averaged over 45-second intervals and confined to times when MAVEN was located outside of the bowshock and in the upstream solar wind \citep{halekas2017}. Using the SWIA measurements, we calculate the solar wind dynamic pressure ($P_{sw}$) using $P_{sw} = \rho_{sw} V_{sw}^2$. 
    

  \subsection{Combining the Mars Express and MAVEN Data}
  \label{combine}
           
\begin{figure}[h]
    \centering
    \includegraphics[width=.8\textwidth]{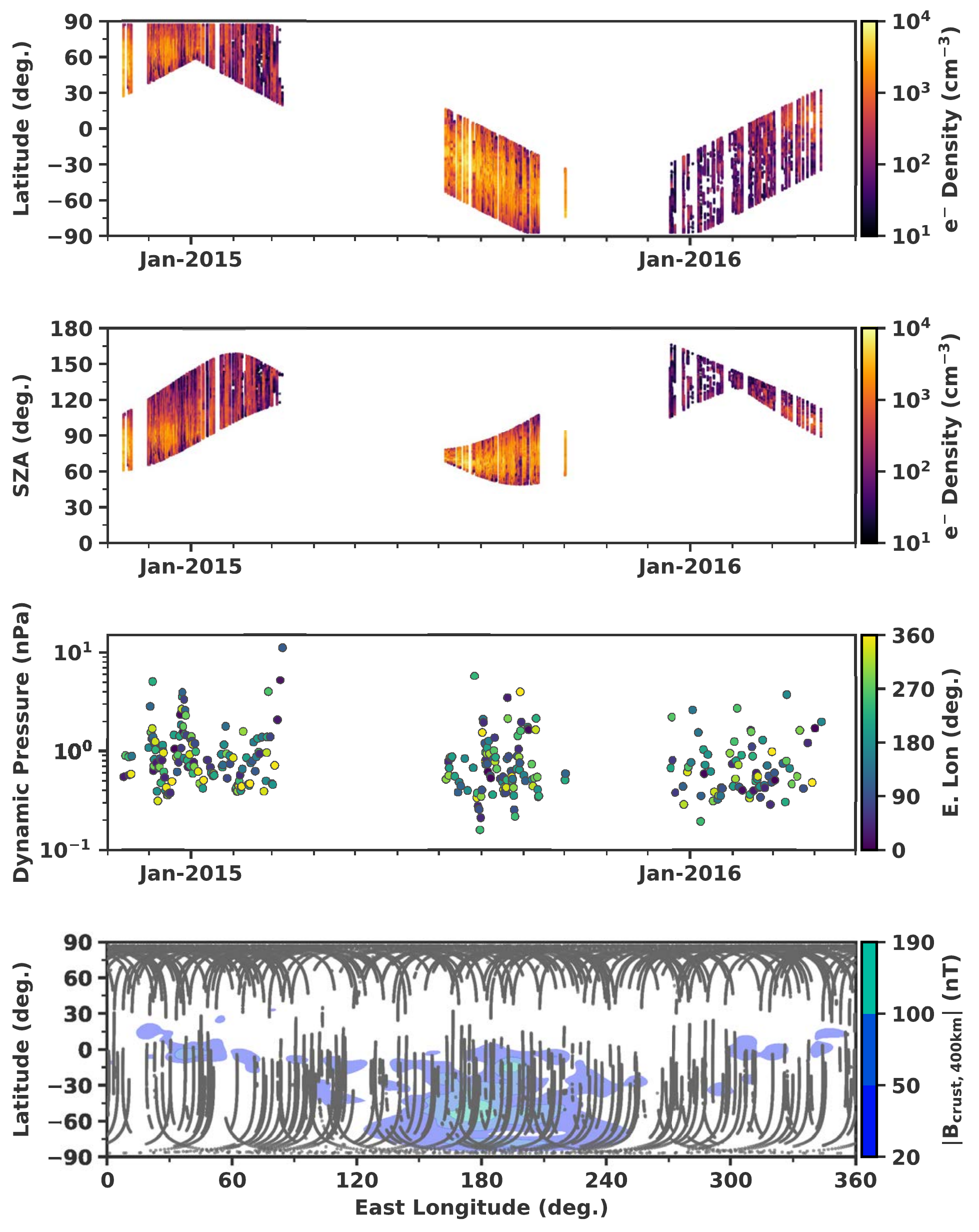}
    \caption{The distribution of the data used in this study. The top two panels show, respectively, the latitudes and solar zenith angles (SZA) of the observations, with the color scale showing the MARSIS electron densities. The third panel shows the solar wind dynamic pressure with colors representing the geographic east longitude of the concurrent MARSIS measurements. The bottom panel shows geographic locations of the MARSIS measurements along with contours of the crustal magnetic field strength at 400 km \citep{morsch2014}.}
    \label{data}
    \end{figure}

    To combine the two datasets, we assign each orbit of MARSIS data a single value of the solar wind dynamic pressure, determined by averaging all of the solar wind measurements that fall within one hour of the first MARSIS measurement within a given orbit. If no solar wind measurements fall within one hour, then we exclude that MARSIS orbit from our analysis. The resultant dataset contains observations from 12 November 2014 to 06 April 2016, and includes more than 30,000 local electron density measurements from 304 MEX orbits. The distribution of the electron densities with respect to geographic latitude and solar zenith angle (SZA) are shown in the top two panels of Figure~\ref{data}. In these panels, the vertical lines span the periapsis segment of MEX during which MARSIS sounded the ionosphere. The midpoint of each vertical line marks the periapsis of MEX, which varied between 320 and 385 km during this time period. Data gaps within individual orbits are somewhat common, particularly on the nightside where densities are often below the MARSIS detection threshold of $\sim$10 cm$^{-3}$.
    
    The third panel in Figure~\ref{data} summarizes the concurrent MAVEN measurements of the upstream solar wind dynamic pressure, which varied between 0.16 - 11.1 nPa. The dynamic pressure during this time period had an average value of 0.98 nPa, and 25\%, 50\%, and 75\% quartile values of 0.48 nPa, 0.70 nPa, and 1.0 nPa, respectively. The data in this panel are colored according to east longitude to show that the dynamic pressures are well distributed with respect to the crustal magnetic fields. The fourth panel in Figure~\ref{data} further demonstrates this by showing the distribution of the measurements with respect to latitude, longitude, and the crustal magnetic field strength at 400 km \citep{morsch2014}. 
    These panels illustrate the data are well-distributed in latitude, longitude, SZA, and solar wind dynamic pressure. 

    The data set contains three distinct observational periods. During the first period, from 12 November 2014 to 08 March 2015, the observations were concentrated at high northern latitudes and covered SZAs from $\sim$60$^{\circ}$ to 160$^{\circ}$. Space weather conditions during this period were somewhat stormy, and included two major solar energetic particle (SEP) events (Dec. 2014 and Feb. 2015), a series of M and X class solar flares, and several coronal mass ejection (CMEs) impacts \citep{jakosky2015b,girazian2017,lee2017}.
    
    During the second period, from 06 July 2015 to 02 October 2015,
    the observations were concentrated at mid/low southern latitudes and covered SZAs from $\sim$45$^{\circ}$ to 105$^{\circ}$. Space weather conditions during this period were more quiescent, but included a handful of minor SEP events, M class solar flares, and weak CME impacts \citep{lee2017}. 
    
    During the third period, from 17 December 2015 to 06 April 2016, the observations were again concentrated at mid/low southern latitudes, but this time covered SZAs from $\sim$90$^{\circ}$ to 170$^{\circ}$. Space weather conditions during this period were somewhat stormy, particularly in January 2016, during which there were two SEP events, M class flares, and CME impacts \citep{lee2017}.

    The extreme ultraviolet (EUV) solar flux was a factor of $\sim$1.5 higher during the first observational period compared to the second and third observational periods \citep{lee2017,girazian2019}. Since dayside electron densities are highly dependent on the EUV flux, this difference must be accounted for when analyzing dayside data from the different time periods. In particular, topside electron densities are generally larger when the EUV flux is higher \citep{duru2019}.
     
    In the next section, we use this combined data set to show how the solar wind dynamic pressure affects electron densities and induced magnetic fields in the topside ionosphere. We also investigate how crustal magnetic fields influence the response of the topside ionosphere to variations in the solar wind dynamic pressure.


    
    
  \section{Effects of Solar Wind Dynamic Pressure}
  \label{pressure}
    
  \subsection{Depletion of the Topside Ionosphere}
   \label{deplete}
  
	To investigate how the topside ionosphere responds to changes in solar wind dynamic pressure, we have computed median electron density profiles at dayside solar zenith angles between 40$^{\circ}$ and 70$^{\circ}$ after separating the data into 25-km altitude bins and into low ($<$0.5 nPa); typical (0.5-0.8 nPa); high (0.8-1.1 nPa); and extreme ($>$1.1 nPa) solar wind dynamic pressure bins. The median density in each bin was computed only if the bin contained at least 30 measurements. The profiles were derived using data from the second and third time periods to ensure that the solar EUV flux was roughly constant throughout the observations (Sec.~\ref{combine}). Additionally, data from strong crustal field regions were removed prior to averaging, with strong crustal fields being defined as those greater than 10 nT at an altitude of 400 km \citep{morsch2014}. 
	
\begin{figure}[h]
    \centering
    \includegraphics[width=1.0\textwidth]{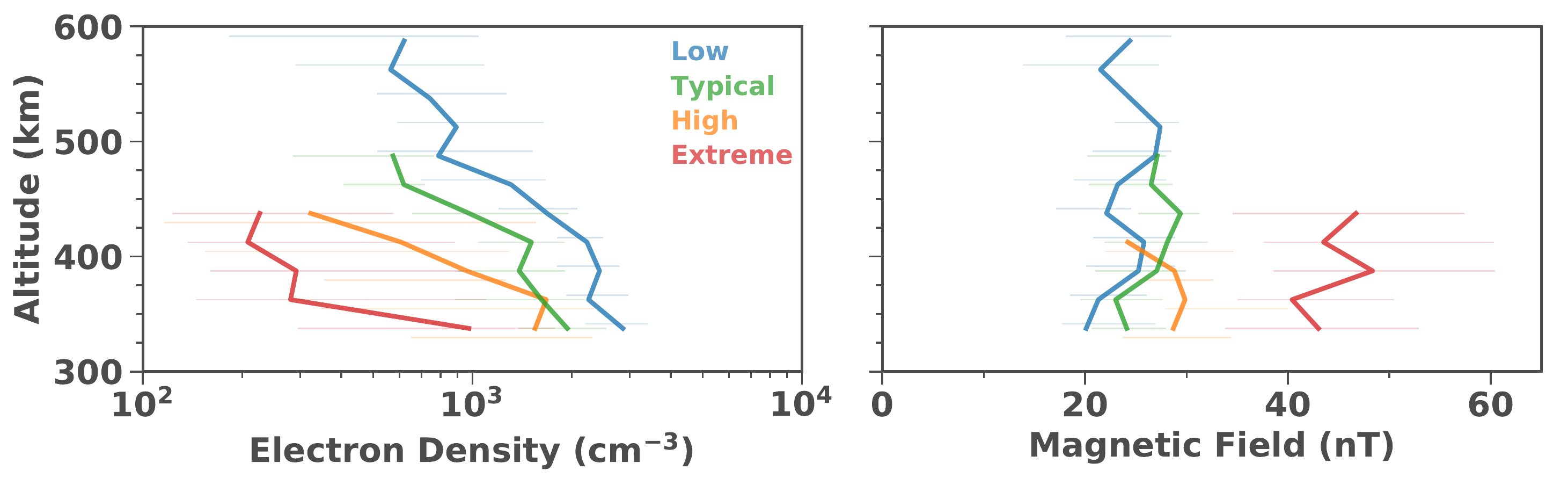}
    \caption{As the solar wind dynamic pressure increases, the topside ionosphere is depleted of plasma (left), and the induced magnetic field strength increases (right). The colors in each panel mark median profiles computed during low, typical, high, and extreme solar wind dynamic pressures as defined in the text. The error bars are the 25\% and 75\% density quartiles in each 25 km altitude bin and are offset in altitude for visual clarity. The data are from dayside solar zenith angles between 40$^{\circ}$ and 70$^{\circ}$ and weak crustal field regions.}
    \label{altprofiles}
    \end{figure}

	The resultant median profiles are shown in the left panel of Figure~\ref{altprofiles}. Note that some of the abrupt structures in the median profiles are a consequence of the data having nonuniform altitude sampling, as there are more measurements at lower altitudes than at higher altitudes. The number of measurements in each averaging bin ranges from several hundreds at the lowest altitudes, to thirty at the highest altitudes, with each bin containing measurements from at least six unique MEX orbits.
	
	The density profiles in Figure~\ref{altprofiles} show that electron densities in the topside ionosphere are strongly affected by the solar wind dynamic pressure. In each successive dynamic pressure bin, the densities become smaller and, at most altitudes, there is clear separation between the different median density profiles. The profiles show that increases in the solar wind dynamic pressure result in the topside ionosphere being depleted of plasma. From low to extreme dynamic pressure conditions, electron densities at $\sim$400 km are depleted by a factor of $\sim$10. 
	
	The right panel in Figure~\ref{altprofiles} shows median profiles of the measured magnetic field strength computed in the same way. There is a clear trend of increasing magnetic field strength with increasing solar wind dynamic pressure, particularly below 400 km. This trend is consistent with previous observations \citep{crider2003,edberg2009}, and matches the expectation that the induced magnetic field strength must increase to maintain pressure balance across the Martian plasma environment \citep{luhmann1987,dubinin2008a,dubinin2008b}.

  \subsection{The Role of Crustal Magnetic Fields}

\begin{figure}[h]
    \centering
   \includegraphics[width=1.0\textwidth]{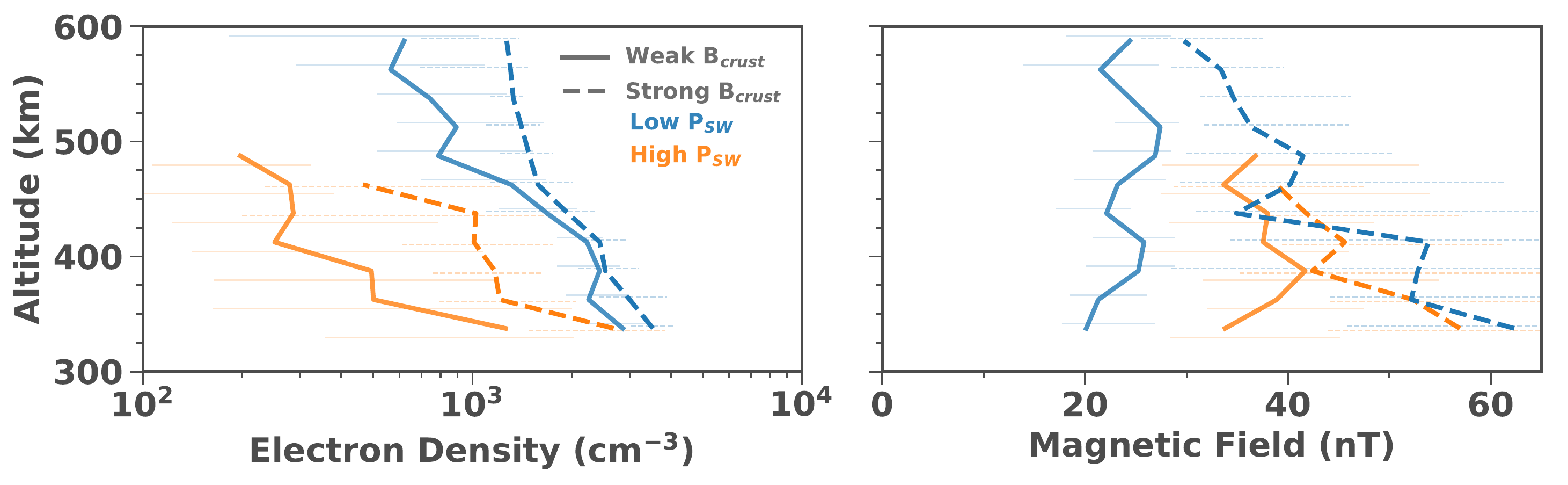}
    \caption{Comparing the response of the topside ionosphere to variations in solar wind dynamic pressure near weak and strong crustal magnetic fields. The left panel shows median electron density profiles and the right panel shows median magnetic field profiles. Solid lines identify profiles from weak crustal field regions ($|B_{\mathrm{crust, 400 km}}|$ $<$ 10 nT) while dashed lines identify profiles from strong crustal field regions ($|B_{\mathrm{crust, 400 km}}|$ $>$ 20 nT). The colors identify profiles from low solar wind dynamic pressure (blue, $<$0.5 nPa) and high solar wind dynamic pressure (orange, $>$0.8 nPa). The error bars are the 25\% and 75\% density quartiles in each 25 km altitude bin and are offset in altitude for visual clarity. The data are from dayside solar zenith angles between 40$^{\circ}$ and 70$^{\circ}$.}
    \label{crust}
    \end{figure}

    To investigate how crustal magnetic fields influence the ionospheric response to variations in solar wind dynamic pressure, we have computed median electron density profiles using a similar procedure as in Section~\ref{deplete}. This time, though, we separated the data into low ($<$0.5 nPa) and high ($>$0.8 nPa) solar wind dynamic pressure bins, and then into weak and strong crustal magnetic fields bins. The crustal field bins are based on the field strength at 400 km \citep{morsch2014}, with weak fields defined as being less than 10 nT, and strong fields defined as being greater than 20 nT (Figure~\ref{data}). The resultant profiles are shown in the left panel of Figure~\ref{crust}. 
    
    By comparing density profiles from weak (dashed) and strong (solid) crustal field regions, it can be seen that higher electron densities are observed near strong crustal fields for both solar wind dynamic pressure cases. This result is consistent with previously reported observations of increased plasma densities in the topside ionosphere near strong crustal fields \citep{andrews2015,nemec2016,flynn2017,nemec2019,withers2019}.
	
    Comparing the two density profiles from weak crustal field regions (solid profiles) shows that they differ in the same way as the profiles shown in Figure~\ref{altprofiles}, once again indicating that increases in the solar wind dynamic pressure result in a depleted topside ionosphere. A similar difference is also observed when comparing the two density profiles from strong crustal field regions (dashed profiles), indicating that the depletion of the topside ionosphere during high solar wind dynamic pressure is also observed in strong crustal field regions.
		
    The right panel in Figure~\ref{crust} shows the median profiles of the magnetic field strength computed from the same orbits. Near weak crustal fields (solid lines), the field strength increases during high solar wind dynamic pressure, consistent with what is shown in Figure~\ref{altprofiles}, with the induced magnetic field strength increasing during high solar wind dynamic pressures. In contrast, near strong crustal fields, there is no significant difference in the magnetic field profiles from both low and high solar wind dynamic pressures, as expected for regions where the crustal magnetic field strength greatly exceeds the induced magnetic field strength \citep{brain2003}. 
    

 \subsection{Influence on Nightside Plasma Sources}
  
	To investigate the global response of the topside ionosphere to variations in solar wind dynamic pressure, Figure~\ref{pressure} compares electron densities during low ($<$ 0.5 nPa) and high ($>$0.7 nPa) solar wind dynamic pressures using all data from locations where the crustal magnetic field strength at 400 km was less than 10 nT \citep{morsch2014}. The electron densities are plotted as a function of altitude and SZA after binning the data into bins of size 25 km in the Mars-Solar-Orbital (MSO) $X$ direction, and 25 km in the MSO $\rho = (Y^2 + Z^2)^{1/2}$ direction.

	From this comparison, it is readily observed that the topside ionosphere is depleted of plasma at all SZAs during times of high solar wind dynamic pressure. This trend is observed deep into the nightside where the ionosphere, at these high altitudes, is populated by combination of electron impact ionization and plasma that has been transported from the dayside \citep{ma2004,chaufray2014,cui2015,girazian2017a,adams2018}. This result shows that transport, impact ionization, or both, are reduced during times of high solar wind dynamic pressure. 
	
	

 \subsection{Impact on Globally Induced Magnetic Fields}
  
\begin{figure}[h]
    \centering
    \includegraphics[width=0.5\textwidth]{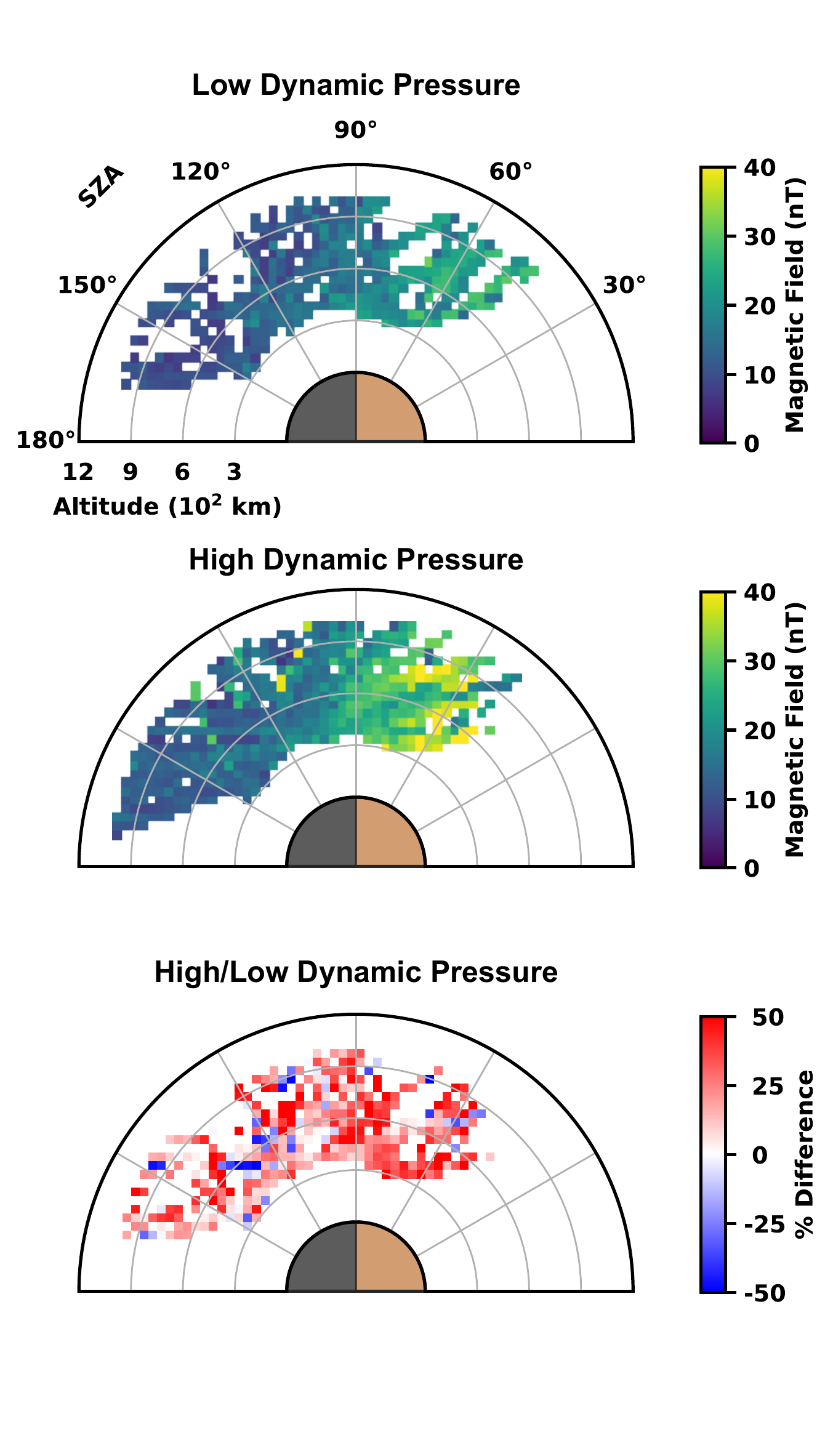}
    \caption{Magnetic field strength increases at all solar zenith angles (SZAs) during times of high solar wind dynamic pressure. The top two panels show magnetic field strength as a function of SZA and altitude for low (top) and high (middle) solar wind dynamic pressures. The bottom panel shows the percent difference between the two panels with positive values (red colors) indicating that the magnetic field is stronger during high dynamic pressure. }
    \label{bfield}
    \end{figure}

    To investigate the global response of the induced magnetic field to variations in solar wind dynamic pressure, Figure~\ref{bfield} compares the measured magnetic fields during low ($<$ 0.5 nPa) and high ($>$0.7 nPa) solar wind dynamic pressures using all the data from locations where the crustal magnetic field strength at 400 km was less than 10 nT \citep{morsch2014}. The magnetic fields are plotted in the same way as the electron densities in Figure~\ref{pressure}.
  
	From this comparison, it is readily observed that the induced magnetic field strength increases at all SZAs during times of high solar wind dynamic pressure. This trend is observed deep into the nightside, indicating that the induced magnetic field is enhanced globally during times of high solar wind dynamic pressure. This finding is consistent with previously reported observations from the Mars Global Surveyor magnetometer \citep{ferguson2005}, and from MARSIS observations during a CME impact \citep{harada2018}.

  \section{Discussion}

    Using MAVEN measurements of the upstream solar wind, and MEX measurements of ionospheric electron densities, we have investigated how the topside ionosphere of Mars is affected by variations in solar wind dynamic pressure. We find that, at all observed SZAs, the topside ionosphere is depleted of plasma during times of high solar wind dynamic pressure, which is broadly consistent with previous case studies of the ionosphere during impulsive space weather events (Section~\ref{intro}), and the statistical study by \citet{dubinin2018}. We further find that induced magnetic field strengths increase at all SZAs during times of high solar wind dynamic pressure, which is consistent with observations from the Mars Global Surveyor magnetometer of stronger draped magnetic fields on the nightside during times of high solar wind dynamic pressure \citep{ferguson2005}.
    
    Our results point to new aspects in regards to how the solar wind dynamic pressure affects the topside ionosphere of Mars. First, we find that topside plasma densities are depleted during times of high solar wind dynamic pressure near both weak and strong crustal magnetic fields. Although electron densities are consistently elevated near strong crustal fields \citep{andrews2015,nemec2016,flynn2017,nemec2019,withers2019}, they are still significantly depleted during times of high solar wind dynamic pressure. Crustal magnetic fields, thus, do not completely shield the ionosphere from the influence of the solar wind.
    
    
    Second, we find that plasma densities at high SZAs on the nightside are depleted during times of high solar wind dynamic pressure. This demonstrates that, during high solar wind dynamic pressure, the production of plasma in the high-altitude nightside ionosphere is reduced. Nightside plasma at these altitudes can be sourced by electron impact ionization or day-to-night plasma transport. Evidence suggests, however, that electron impact ionization is the primary source of plasma at low altitudes ($<200$ km), and day-to-night transport is the primary source at high altitudes \citep{girazian2017a,adams2018,cao2019,cui2019}. Interpretation of this result, then, might be explained by analogy with the ionosphere of Venus \citep{cravens1982,miller1984}: during times of high solar wind dynamic pressure, there is a smaller reservoir of dayside plasma that can be transported to maintain the high-altitude nightside ionosphere.

    

   

    
    Interpretation of our findings is somewhat limited because MARSIS cannot measure vector components of the magnetic field, nor bulk properties of the topside ionospheric plasma. Nonetheless, it is well known that high solar wind dynamic pressures compress the Martian plasma environment \citep{crider2003,edberg2009,opgen2013,ma2014,halekas2017,halekas2018}, allowing for stronger, and deeper penetrating, draped and open magnetic fields \citep{crider2003,crider2005,jakosky2015b,xu2018,fowler2019,xu2019}. In such a scenario, the topside ionosphere can be depleted due to higher ion escape rates, driven by stronger magnetic tension and pressure gradient forces \citep{cravens2017,halekas2017a,wu2019}, increased pickup ion escape, and enhanced ion outflow.

  \section{Conclusions}
    
    In conclusion, the solar wind dynamic pressure significantly affects electron densities in the topside ionosphere of Mars. Increases in the solar wind dynamic pressure result in stronger induced magnetic fields, and a depleted topside ionosphere at all SZAs, near both weak and strong crustal magnetic fields. Interpretation of these results suggests that high solar wind dynamic pressures result in ionospheric compression, increased ion escape, and reduced day-to-night plasma flow. Future comparisons of our results with MHD models, and with MAVEN plasma and magnetic field data, will improve our understanding of the physical processes that lead to the solar wind dynamic pressure having a significant influence on the structure of the topside ionosphere of Mars.

\acknowledgments
    The authors give special thanks to Chris Piker for helping with data acquisition and plotting techniques, and to Joe Groene for helping with SPICE coordinate conversions. ZG thanks Bill Kurth, Ali Sulaiman, and Masafumi Imai for their useful inputs regarding this work. The research presented here was supported by NASA through Contract No. 1560641 with the Jet Propulsion Laboratory. The MAVEN mission is supported through NASA headquarters. The MAVEN and MARSIS data used in this publication are publicly available and can be downloaded from the NASA Planetary Data System (PDS).

.


%
%

\bibliography{References_masterPW.bib}

\begin{thebibliography}{}

\bibitem [\protect \citeauthoryear {%
Adams%
\ \protect \BOthers {.}}{%
Adams%
\ \protect \BOthers {.}}{%
{\protect \APACyear {2018}}%
}]{%
adams2018}
\APACinsertmetastar {%
adams2018}%
\begin{APACrefauthors}%
Adams, D.%
, Xu, S.%
, Mitchell, D\BPBI L.%
, Lillis, R\BPBI J.%
, Fillingim, M.%
, Andersson, L.%
\BDBL {}Mazelle, C.%
\end{APACrefauthors}%
\unskip\
\newblock
\APACrefYearMonthDay{2018}{}{}.
\newblock
{\BBOQ}\APACrefatitle {Using Magnetic Topology to Probe the Sources of {M}ars'
  Nightside Ionosphere} {Using magnetic topology to probe the sources of
  {M}ars' nightside ionosphere}.{\BBCQ}
\newblock
\APACjournalVolNumPages{Geophys. Res. Lett.}{45}{22}{12,190-12,197}.
\newblock
\begin{APACrefURL}
  \url{https://agupubs.onlinelibrary.wiley.com/doi/abs/10.1029/2018GL080629}
  \end{APACrefURL}
\newblock
\begin{APACrefDOI} \doi{10.1029/2018GL080629} \end{APACrefDOI}
\PrintBackRefs{\CurrentBib}

\bibitem [\protect \citeauthoryear {%
{Akalin}%
\ \protect \BOthers {.}}{%
{Akalin}%
\ \protect \BOthers {.}}{%
{\protect \APACyear {2010}}%
}]{%
akalin2010}
\APACinsertmetastar {%
akalin2010}%
\begin{APACrefauthors}%
{Akalin}, F.%
, {Morgan}, D\BPBI D.%
, {Gurnett}, D\BPBI A.%
, {Kirchner}, D\BPBI L.%
, {Brain}, D\BPBI A.%
, {Modolo}, R.%
\BDBL {}{Espley}, J\BPBI R.%
\end{APACrefauthors}%
\unskip\
\newblock
\APACrefYearMonthDay{2010}{}{}.
\newblock
{\BBOQ}\APACrefatitle {{Dayside induced magnetic field in the ionosphere of
  {M}ars}} {{Dayside induced magnetic field in the ionosphere of
  {M}ars}}.{\BBCQ}
\newblock
\APACjournalVolNumPages{Icarus}{206}{1}{104-111}.
\newblock
\begin{APACrefDOI} \doi{10.1016/j.icarus.2009.03.021} \end{APACrefDOI}
\PrintBackRefs{\CurrentBib}

\bibitem [\protect \citeauthoryear {%
{Andrews}%
\ \protect \BOthers {.}}{%
{Andrews}%
\ \protect \BOthers {.}}{%
{\protect \APACyear {2015}}%
}]{%
andrews2015}
\APACinsertmetastar {%
andrews2015}%
\begin{APACrefauthors}%
{Andrews}, D\BPBI J.%
, {Andersson}, L.%
, {Delory}, G\BPBI T.%
, {Ergun}, R\BPBI E.%
, {Eriksson}, A\BPBI I.%
, {Fowler}, C\BPBI M.%
\BDBL {}{Jakosky}, B\BPBI M.%
\end{APACrefauthors}%
\unskip\
\newblock
\APACrefYearMonthDay{2015}{}{}.
\newblock
{\BBOQ}\APACrefatitle {{Ionospheric plasma density variations observed at
  {M}ars by {MAVEN/LPW}}} {{Ionospheric plasma density variations observed at
  {M}ars by {MAVEN/LPW}}}.{\BBCQ}
\newblock
\APACjournalVolNumPages{Geophys. Res. lett.}{42}{}{8862-8869}.
\newblock
\begin{APACrefDOI} \doi{10.1002/2015GL065241} \end{APACrefDOI}
\PrintBackRefs{\CurrentBib}

\bibitem [\protect \citeauthoryear {%
{Brain}%
, {Bagenal}%
, {Acu{\~n}a}%
\BCBL {}\ \BBA {} {Connerney}%
}{%
{Brain}%
\ \protect \BOthers {.}}{%
{\protect \APACyear {2003}}%
}]{%
brain2003}
\APACinsertmetastar {%
brain2003}%
\begin{APACrefauthors}%
{Brain}, D\BPBI A.%
, {Bagenal}, F.%
, {Acu{\~n}a}, M\BPBI H.%
\BCBL {}\ \BBA {} {Connerney}, J\BPBI E\BPBI P.%
\end{APACrefauthors}%
\unskip\
\newblock
\APACrefYearMonthDay{2003}{}{}.
\newblock
{\BBOQ}\APACrefatitle {Martian magnetic morphology: Contributions from the
  solar wind and crust} {Martian magnetic morphology: Contributions from the
  solar wind and crust}.{\BBCQ}
\newblock
\APACjournalVolNumPages{J. Geophys. Res.}{108}{}{1424, 10.1029/2002JA009482}.
\newblock
\begin{APACrefDOI} \doi{10.1029/2002JA009482} \end{APACrefDOI}
\PrintBackRefs{\CurrentBib}

\bibitem [\protect \citeauthoryear {%
{Brain}%
, {Bagenal}%
, {Ma}%
, {Nilsson}%
\BCBL {}\ \BBA {} {Stenberg Wieser}%
}{%
{Brain}%
\ \protect \BOthers {.}}{%
{\protect \APACyear {2016}}%
}]{%
brain2016}
\APACinsertmetastar {%
brain2016}%
\begin{APACrefauthors}%
{Brain}, D\BPBI A.%
, {Bagenal}, F.%
, {Ma}, Y\BHBI J.%
, {Nilsson}, H.%
\BCBL {}\ \BBA {} {Stenberg Wieser}, G.%
\end{APACrefauthors}%
\unskip\
\newblock
\APACrefYearMonthDay{2016}{}{}.
\newblock
{\BBOQ}\APACrefatitle {{Atmospheric escape from unmagnetized bodies}}
  {{Atmospheric escape from unmagnetized bodies}}.{\BBCQ}
\newblock
\APACjournalVolNumPages{J. Geophys. Res.}{121}{}{2364-2385}.
\newblock
\begin{APACrefDOI} \doi{10.1002/2016JE005162} \end{APACrefDOI}
\PrintBackRefs{\CurrentBib}

\bibitem [\protect \citeauthoryear {%
{Brain}%
\ \protect \BOthers {.}}{%
{Brain}%
\ \protect \BOthers {.}}{%
{\protect \APACyear {2005}}%
}]{%
brain2005}
\APACinsertmetastar {%
brain2005}%
\begin{APACrefauthors}%
{Brain}, D\BPBI A.%
, {Halekas}, J\BPBI S.%
, {Lillis}, R.%
, {Mitchell}, D\BPBI L.%
, {Lin}, R\BPBI P.%
\BCBL {}\ \BBA {} {Crider}, D\BPBI H.%
\end{APACrefauthors}%
\unskip\
\newblock
\APACrefYearMonthDay{2005}{}{}.
\newblock
{\BBOQ}\APACrefatitle {{Variability of the altitude of the {M}artian sheath}}
  {{Variability of the altitude of the {M}artian sheath}}.{\BBCQ}
\newblock
\APACjournalVolNumPages{Geophys. Res. Lett.}{32}{18}{L18203}.
\newblock
\begin{APACrefDOI} \doi{10.1029/2005GL023126} \end{APACrefDOI}
\PrintBackRefs{\CurrentBib}

\bibitem [\protect \citeauthoryear {%
Cao%
, Cui%
, Wu%
, Guo%
\BCBL {}\ \BBA {} Wei%
}{%
Cao%
\ \protect \BOthers {.}}{%
{\protect \APACyear {2019}}%
}]{%
cao2019}
\APACinsertmetastar {%
cao2019}%
\begin{APACrefauthors}%
Cao, Y\BHBI T.%
, Cui, J.%
, Wu, X\BHBI S.%
, Guo, J\BHBI P.%
\BCBL {}\ \BBA {} Wei, Y.%
\end{APACrefauthors}%
\unskip\
\newblock
\APACrefYearMonthDay{2019}{}{}.
\newblock
{\BBOQ}\APACrefatitle {Structural variability of the nightside Martian
  ionosphere near the terminator: Implications on plasma sources} {Structural
  variability of the nightside martian ionosphere near the terminator:
  Implications on plasma sources}.{\BBCQ}
\newblock
\APACjournalVolNumPages{J. Geophys. Res.}{0}{ja}{}.
\newblock
\begin{APACrefURL}
  \url{https://agupubs.onlinelibrary.wiley.com/doi/abs/10.1029/2019JE005970}
  \end{APACrefURL}
\newblock
\begin{APACrefDOI} \doi{10.1029/2019JE005970} \end{APACrefDOI}
\PrintBackRefs{\CurrentBib}

\bibitem [\protect \citeauthoryear {%
{Chaufray}%
\ \protect \BOthers {.}}{%
{Chaufray}%
\ \protect \BOthers {.}}{%
{\protect \APACyear {2014}}%
}]{%
chaufray2014}
\APACinsertmetastar {%
chaufray2014}%
\begin{APACrefauthors}%
{Chaufray}, J\BHBI Y.%
, {Gonzalez-Galindo}, F.%
, {Forget}, F.%
, {Lopez-Valverde}, M.%
, {Leblanc}, F.%
, {Modolo}, R.%
\BDBL {}{Witasse}, O.%
\end{APACrefauthors}%
\unskip\
\newblock
\APACrefYearMonthDay{2014}{}{}.
\newblock
{\BBOQ}\APACrefatitle {{Three-dimensional {M}artian ionosphere model: {II}.
  {E}ffect of transport processes due to pressure gradients}}
  {{Three-dimensional {M}artian ionosphere model: {II}. {E}ffect of transport
  processes due to pressure gradients}}.{\BBCQ}
\newblock
\APACjournalVolNumPages{J. Geophys. Res.}{119}{}{1614-1636}.
\newblock
\begin{APACrefDOI} \doi{10.1002/2013JE004551} \end{APACrefDOI}
\PrintBackRefs{\CurrentBib}

\bibitem [\protect \citeauthoryear {%
Chen%
, Cravens%
\BCBL {}\ \BBA {} Nagy%
}{%
Chen%
\ \protect \BOthers {.}}{%
{\protect \APACyear {1978}}%
}]{%
chen1978}
\APACinsertmetastar {%
chen1978}%
\begin{APACrefauthors}%
Chen, R\BPBI H.%
, Cravens, T\BPBI E.%
\BCBL {}\ \BBA {} Nagy, A\BPBI F.%
\end{APACrefauthors}%
\unskip\
\newblock
\APACrefYearMonthDay{1978}{}{}.
\newblock
{\BBOQ}\APACrefatitle {The martian ionosphere in light of the {V}iking
  observations} {The martian ionosphere in light of the {V}iking
  observations}.{\BBCQ}
\newblock
\APACjournalVolNumPages{J. Geophys. Res.}{83}{}{3871-3876}.
\newblock
\begin{APACrefDOI} \doi{10.1029/JA083iA08p03871} \end{APACrefDOI}
\PrintBackRefs{\CurrentBib}

\bibitem [\protect \citeauthoryear {%
{Collinson}%
\ \protect \BOthers {.}}{%
{Collinson}%
\ \protect \BOthers {.}}{%
{\protect \APACyear {2018}}%
}]{%
collinson2018}
\APACinsertmetastar {%
collinson2018}%
\begin{APACrefauthors}%
{Collinson}, G.%
, {Wilson}, L\BPBI B.%
, {Omidi}, N.%
, {Sibeck}, D.%
, {Espley}, J.%
, {Fowler}, C\BPBI M.%
\BDBL {}{Jakosky}, B.%
\end{APACrefauthors}%
\unskip\
\newblock
\APACrefYearMonthDay{2018}{}{}.
\newblock
{\BBOQ}\APACrefatitle {{Solar Wind Induced Waves in the Skies of {M}ars:
  {I}onospheric Compression, Energization, and Escape Resulting From the Impact
  of Ultralow Frequency Magnetosonic Waves Generated Upstream of the Martian
  Bow Shock}} {{Solar Wind Induced Waves in the Skies of {M}ars: {I}onospheric
  Compression, Energization, and Escape Resulting From the Impact of Ultralow
  Frequency Magnetosonic Waves Generated Upstream of the Martian Bow
  Shock}}.{\BBCQ}
\newblock
\APACjournalVolNumPages{J. Geophys. Res.}{123}{9}{7241-7256}.
\newblock
\begin{APACrefDOI} \doi{10.1029/2018JA025414} \end{APACrefDOI}
\PrintBackRefs{\CurrentBib}

\bibitem [\protect \citeauthoryear {%
{Cravens}%
\ \protect \BOthers {.}}{%
{Cravens}%
\ \protect \BOthers {.}}{%
{\protect \APACyear {1982}}%
}]{%
cravens1982}
\APACinsertmetastar {%
cravens1982}%
\begin{APACrefauthors}%
{Cravens}, T\BPBI E.%
, {Brace}, L\BPBI H.%
, {Taylor}, H\BPBI A.%
, {Russell}, C\BPBI T.%
, {Knudsen}, W\BPBI L.%
, {Miller}, K\BPBI L.%
\BDBL {}{Nagy}, A\BPBI F.%
\end{APACrefauthors}%
\unskip\
\newblock
\APACrefYearMonthDay{1982}{}{}.
\newblock
{\BBOQ}\APACrefatitle {{Disappearing ionospheres on the nightside of {V}enus}}
  {{Disappearing ionospheres on the nightside of {V}enus}}.{\BBCQ}
\newblock
\APACjournalVolNumPages{Icarus}{51}{2}{271-282}.
\newblock
\begin{APACrefDOI} \doi{10.1016/0019-1035(82)90083-5} \end{APACrefDOI}
\PrintBackRefs{\CurrentBib}

\bibitem [\protect \citeauthoryear {%
{Cravens}%
\ \protect \BOthers {.}}{%
{Cravens}%
\ \protect \BOthers {.}}{%
{\protect \APACyear {2017}}%
}]{%
cravens2017}
\APACinsertmetastar {%
cravens2017}%
\begin{APACrefauthors}%
{Cravens}, T\BPBI E.%
, {Hamil}, O.%
, {Houston}, S.%
, {Bougher}, S.%
, {Ma}, Y.%
, {Brain}, D.%
\BCBL {}\ \BBA {} {Ledvina}, S.%
\end{APACrefauthors}%
\unskip\
\newblock
\APACrefYearMonthDay{2017}{}{}.
\newblock
{\BBOQ}\APACrefatitle {{Estimates of Ionospheric Transport and Ion Loss at
  {M}ars}} {{Estimates of Ionospheric Transport and Ion Loss at
  {M}ars}}.{\BBCQ}
\newblock
\APACjournalVolNumPages{J. Geophys. Res.}{122}{10}{10,626-10,637}.
\newblock
\begin{APACrefDOI} \doi{10.1002/2017JA024582} \end{APACrefDOI}
\PrintBackRefs{\CurrentBib}

\bibitem [\protect \citeauthoryear {%
{Crider}%
\ \protect \BOthers {.}}{%
{Crider}%
\ \protect \BOthers {.}}{%
{\protect \APACyear {2005}}%
}]{%
crider2005}
\APACinsertmetastar {%
crider2005}%
\begin{APACrefauthors}%
{Crider}, D\BPBI H.%
, {Espley}, J.%
, {Brain}, D\BPBI A.%
, {Mitchell}, D\BPBI L.%
, {Connerney}, J\BPBI E\BPBI P.%
\BCBL {}\ \BBA {} {Acu{\~n}A}, M\BPBI H.%
\end{APACrefauthors}%
\unskip\
\newblock
\APACrefYearMonthDay{2005}{}{}.
\newblock
{\BBOQ}\APACrefatitle {{{M}ars {G}lobal {S}urveyor observations of the
  {H}alloween 2003 solar superstorm's encounter with Mars}} {{{M}ars {G}lobal
  {S}urveyor observations of the {H}alloween 2003 solar superstorm's encounter
  with Mars}}.{\BBCQ}
\newblock
\APACjournalVolNumPages{J. Geophys. Res.}{110}{A9}{A09S21}.
\newblock
\begin{APACrefDOI} \doi{10.1029/2004JA010881} \end{APACrefDOI}
\PrintBackRefs{\CurrentBib}

\bibitem [\protect \citeauthoryear {%
{Crider}%
\ \protect \BOthers {.}}{%
{Crider}%
\ \protect \BOthers {.}}{%
{\protect \APACyear {2003}}%
}]{%
crider2003}
\APACinsertmetastar {%
crider2003}%
\begin{APACrefauthors}%
{Crider}, D\BPBI H.%
, {Vignes}, D.%
, {Krymskii}, A\BPBI M.%
, {Breus}, T\BPBI K.%
, {Ness}, N\BPBI F.%
, {Mitchell}, D\BPBI L.%
\BDBL {}{Acu{\~n}a}, M\BPBI H.%
\end{APACrefauthors}%
\unskip\
\newblock
\APACrefYearMonthDay{2003}{}{}.
\newblock
{\BBOQ}\APACrefatitle {A proxy for determining solar wind dynamic pressure at
  {M}ars using {M}ars {G}lobal {S}urveyor data} {A proxy for determining solar
  wind dynamic pressure at {M}ars using {M}ars {G}lobal {S}urveyor
  data}.{\BBCQ}
\newblock
\APACjournalVolNumPages{J. Geophys. Res.}{108}{}{1461, 10.1029/2003JA009875}.
\newblock
\begin{APACrefDOI} \doi{10.1029/2003JA009875} \end{APACrefDOI}
\PrintBackRefs{\CurrentBib}

\bibitem [\protect \citeauthoryear {%
{Cui}%
\ \protect \BOthers {.}}{%
{Cui}%
\ \protect \BOthers {.}}{%
{\protect \APACyear {2019}}%
}]{%
cui2019}
\APACinsertmetastar {%
cui2019}%
\begin{APACrefauthors}%
{Cui}, J.%
, {Cao}, Y\BPBI T.%
, {Wu}, X\BPBI S.%
, {Xu}, S\BPBI S.%
, {Yelle}, R\BPBI V.%
, {Stone}, S.%
\BDBL {}{Wei}, Y.%
\end{APACrefauthors}%
\unskip\
\newblock
\APACrefYearMonthDay{2019}{}{}.
\newblock
{\BBOQ}\APACrefatitle {{Evaluating Local Ionization Balance in the Nightside
  Martian Upper Atmosphere during {MAVEN} Deep Dip Campaigns}} {{Evaluating
  Local Ionization Balance in the Nightside Martian Upper Atmosphere during
  {MAVEN} Deep Dip Campaigns}}.{\BBCQ}
\newblock
\APACjournalVolNumPages{Astrophys. J.}{}{}{}.
\PrintBackRefs{\CurrentBib}

\bibitem [\protect \citeauthoryear {%
{Cui}%
, {Galand}%
, {Yelle}%
, {Wei}%
\BCBL {}\ \BBA {} {Zhang}%
}{%
{Cui}%
\ \protect \BOthers {.}}{%
{\protect \APACyear {2015}}%
}]{%
cui2015}
\APACinsertmetastar {%
cui2015}%
\begin{APACrefauthors}%
{Cui}, J.%
, {Galand}, M.%
, {Yelle}, R\BPBI V.%
, {Wei}, Y.%
\BCBL {}\ \BBA {} {Zhang}, S\BHBI J.%
\end{APACrefauthors}%
\unskip\
\newblock
\APACrefYearMonthDay{2015}{}{}.
\newblock
{\BBOQ}\APACrefatitle {{Day-to-night transport in the {M}artian ionosphere:
  {I}mplications from total electron content measurements}} {{Day-to-night
  transport in the {M}artian ionosphere: {I}mplications from total electron
  content measurements}}.{\BBCQ}
\newblock
\APACjournalVolNumPages{J. Geophys. Res.}{120}{}{2333-2346}.
\newblock
\begin{APACrefDOI} \doi{10.1002/2014JA020788} \end{APACrefDOI}
\PrintBackRefs{\CurrentBib}

\bibitem [\protect \citeauthoryear {%
{Dong}%
\ \protect \BOthers {.}}{%
{Dong}%
\ \protect \BOthers {.}}{%
{\protect \APACyear {2015}}%
}]{%
dong2015}
\APACinsertmetastar {%
dong2015}%
\begin{APACrefauthors}%
{Dong}, Y.%
, {Fang}, X.%
, {Brain}, D\BPBI A.%
, {McFadden}, J\BPBI P.%
, {Halekas}, J\BPBI S.%
, {Connerney}, J\BPBI E.%
\BDBL {}{Jakosky}, B\BPBI M.%
\end{APACrefauthors}%
\unskip\
\newblock
\APACrefYearMonthDay{2015}{}{}.
\newblock
{\BBOQ}\APACrefatitle {{Strong plume fluxes at {M}ars observed by {MAVEN}: {A}n
  important planetary ion escape channel}} {{Strong plume fluxes at {M}ars
  observed by {MAVEN}: {A}n important planetary ion escape channel}}.{\BBCQ}
\newblock
\APACjournalVolNumPages{Geophys. Res. Lett.}{42}{21}{8942-8950}.
\newblock
\begin{APACrefDOI} \doi{10.1002/2015GL065346} \end{APACrefDOI}
\PrintBackRefs{\CurrentBib}

\bibitem [\protect \citeauthoryear {%
{Dong}%
\ \protect \BOthers {.}}{%
{Dong}%
\ \protect \BOthers {.}}{%
{\protect \APACyear {2017}}%
}]{%
dong2017}
\APACinsertmetastar {%
dong2017}%
\begin{APACrefauthors}%
{Dong}, Y.%
, {Fang}, X.%
, {Brain}, D\BPBI A.%
, {McFadden}, J\BPBI P.%
, {Halekas}, J\BPBI S.%
, {Connerney}, J\BPBI E\BPBI P.%
\BDBL {}{Jakosky}, B\BPBI M.%
\end{APACrefauthors}%
\unskip\
\newblock
\APACrefYearMonthDay{2017}{}{}.
\newblock
{\BBOQ}\APACrefatitle {{Seasonal variability of {M}artian ion escape through
  the plume and tail from {MAVEN} observations}} {{Seasonal variability of
  {M}artian ion escape through the plume and tail from {MAVEN}
  observations}}.{\BBCQ}
\newblock
\APACjournalVolNumPages{J. Geophys. Res.}{122}{}{4009-4022}.
\newblock
\begin{APACrefDOI} \doi{10.1002/2016JA023517} \end{APACrefDOI}
\PrintBackRefs{\CurrentBib}

\bibitem [\protect \citeauthoryear {%
{Dubinin}%
\ \protect \BOthers {.}}{%
{Dubinin}%
\ \protect \BOthers {.}}{%
{\protect \APACyear {2011}}%
}]{%
dubinin2011}
\APACinsertmetastar {%
dubinin2011}%
\begin{APACrefauthors}%
{Dubinin}, E.%
, {Fraenz}, M.%
, {Fedorov}, A.%
, {Lundin}, R.%
, {Edberg}, N.%
, {Duru}, F.%
\BCBL {}\ \BBA {} {Vaisberg}, O.%
\end{APACrefauthors}%
\unskip\
\newblock
\APACrefYearMonthDay{2011}{}{}.
\newblock
{\BBOQ}\APACrefatitle {{Ion {E}nergization and {E}scape on {M}ars and {V}enus}}
  {{Ion {E}nergization and {E}scape on {M}ars and {V}enus}}.{\BBCQ}
\newblock
\APACjournalVolNumPages{Space Sci. Rev.}{162}{}{173-211}.
\newblock
\begin{APACrefDOI} \doi{10.1007/s11214-011-9831-7} \end{APACrefDOI}
\PrintBackRefs{\CurrentBib}

\bibitem [\protect \citeauthoryear {%
{Dubinin}%
\ \protect \BOthers {.}}{%
{Dubinin}%
\ \protect \BOthers {.}}{%
{\protect \APACyear {2017}}%
}]{%
dubinin2017b}
\APACinsertmetastar {%
dubinin2017b}%
\begin{APACrefauthors}%
{Dubinin}, E.%
, {Fraenz}, M.%
, {P{\"a}tzold}, M.%
, {McFadden}, J.%
, {Halekas}, J\BPBI S.%
, {DiBraccio}, G\BPBI A.%
\BDBL {}{Zelenyi}, L.%
\end{APACrefauthors}%
\unskip\
\newblock
\APACrefYearMonthDay{2017}{}{}.
\newblock
{\BBOQ}\APACrefatitle {{The Effect of Solar Wind Variations on the Escape of
  Oxygen Ions From {M}ars Through Different Channels: MAVEN Observations}}
  {{The Effect of Solar Wind Variations on the Escape of Oxygen Ions From
  {M}ars Through Different Channels: MAVEN Observations}}.{\BBCQ}
\newblock
\APACjournalVolNumPages{J. Geophys. Res.}{122}{11}{11,285-11,301}.
\newblock
\begin{APACrefDOI} \doi{10.1002/2017JA024741} \end{APACrefDOI}
\PrintBackRefs{\CurrentBib}

\bibitem [\protect \citeauthoryear {%
Dubinin%
\ \protect \BOthers {.}}{%
Dubinin%
\ \protect \BOthers {.}}{%
{\protect \APACyear {2018}}%
}]{%
dubinin2018}
\APACinsertmetastar {%
dubinin2018}%
\begin{APACrefauthors}%
Dubinin, E.%
, Fraenz, M.%
, Pätzold, M.%
, McFadden, J.%
, Halekas, J.%
, Connerney, J.%
\BDBL {}Zelenyi, L.%
\end{APACrefauthors}%
\unskip\
\newblock
\APACrefYearMonthDay{2018}{}{}.
\newblock
{\BBOQ}\APACrefatitle {Martian ionosphere observed by {MAVEN}. 3. {I}nfluence
  of solar wind and {IMF} on upper ionosphere} {Martian ionosphere observed by
  {MAVEN}. 3. {I}nfluence of solar wind and {IMF} on upper ionosphere}.{\BBCQ}
\newblock
\APACjournalVolNumPages{Planet. Space Sci.}{}{}{-}.
\newblock
\begin{APACrefURL}
  \url{https://www.sciencedirect.com/science/article/pii/S0032063317304361}
  \end{APACrefURL}
\newblock
\begin{APACrefDOI} \doi{https://doi.org/10.1016/j.pss.2018.03.016}
  \end{APACrefDOI}
\PrintBackRefs{\CurrentBib}

\bibitem [\protect \citeauthoryear {%
{Dubinin}%
\ \protect \BOthers {.}}{%
{Dubinin}%
\ \protect \BOthers {.}}{%
{\protect \APACyear {2009}}%
}]{%
dubinin2009}
\APACinsertmetastar {%
dubinin2009}%
\begin{APACrefauthors}%
{Dubinin}, E.%
, {Fraenz}, M.%
, {Woch}, J.%
, {Duru}, F.%
, {Gurnett}, D.%
, {Modolo}, R.%
\BDBL {}{Lundin}, R.%
\end{APACrefauthors}%
\unskip\
\newblock
\APACrefYearMonthDay{2009}{}{}.
\newblock
{\BBOQ}\APACrefatitle {{Ionospheric storms on {M}ars: {I}mpact of the
  corotating interaction region}} {{Ionospheric storms on {M}ars: {I}mpact of
  the corotating interaction region}}.{\BBCQ}
\newblock
\APACjournalVolNumPages{Geophys. Res. Lett.}{36}{1}{L01105}.
\newblock
\begin{APACrefDOI} \doi{10.1029/2008GL036559} \end{APACrefDOI}
\PrintBackRefs{\CurrentBib}

\bibitem [\protect \citeauthoryear {%
{Dubinin}%
, {Modolo}%
, {Fraenz}%
, {Woch}%
, {Chanteur}%
\BCBL {}\ \protect \BOthers {.}}{%
{Dubinin}%
, {Modolo}%
, {Fraenz}%
, {Woch}%
, {Chanteur}%
\BCBL {}\ \protect \BOthers {.}}{%
{\protect \APACyear {2008}}%
}]{%
dubinin2008a}
\APACinsertmetastar {%
dubinin2008a}%
\begin{APACrefauthors}%
{Dubinin}, E.%
, {Modolo}, R.%
, {Fraenz}, M.%
, {Woch}, J.%
, {Chanteur}, G.%
, {Duru}, F.%
\BDBL {}{Picardi}, G.%
\end{APACrefauthors}%
\unskip\
\newblock
\APACrefYearMonthDay{2008}{}{}.
\newblock
{\BBOQ}\APACrefatitle {Plasma environment of {M}ars as observed by simultaneous
  {MEX-ASPERA-3} and {MEX-MARSIS} observations} {Plasma environment of {M}ars
  as observed by simultaneous {MEX-ASPERA-3} and {MEX-MARSIS}
  observations}.{\BBCQ}
\newblock
\APACjournalVolNumPages{J. Geophys. Res.}{113}{}{A10217, 10.1029/2008JA013355}.
\newblock
\begin{APACrefDOI} \doi{10.1029/2008JA013355} \end{APACrefDOI}
\PrintBackRefs{\CurrentBib}

\bibitem [\protect \citeauthoryear {%
{Dubinin}%
, {Modolo}%
, {Fraenz}%
, {Woch}%
, {Duru}%
\BCBL {}\ \protect \BOthers {.}}{%
{Dubinin}%
, {Modolo}%
, {Fraenz}%
, {Woch}%
, {Duru}%
\BCBL {}\ \protect \BOthers {.}}{%
{\protect \APACyear {2008}}%
}]{%
dubinin2008b}
\APACinsertmetastar {%
dubinin2008b}%
\begin{APACrefauthors}%
{Dubinin}, E.%
, {Modolo}, R.%
, {Fraenz}, M.%
, {Woch}, J.%
, {Duru}, F.%
, {Akalin}, F.%
\BDBL {}{Picardi}, G.%
\end{APACrefauthors}%
\unskip\
\newblock
\APACrefYearMonthDay{2008}{}{}.
\newblock
{\BBOQ}\APACrefatitle {{Structure and dynamics of the solar wind/ionosphere
  interface on Mars: MEX-ASPERA-3 and MEX-MARSIS observations}} {{Structure and
  dynamics of the solar wind/ionosphere interface on Mars: MEX-ASPERA-3 and
  MEX-MARSIS observations}}.{\BBCQ}
\newblock
\APACjournalVolNumPages{Geophys. Res. Lett.}{35}{11}{L11103}.
\newblock
\begin{APACrefDOI} \doi{10.1029/2008GL033730} \end{APACrefDOI}
\PrintBackRefs{\CurrentBib}

\bibitem [\protect \citeauthoryear {%
{Duru}%
\ \protect \BOthers {.}}{%
{Duru}%
\ \protect \BOthers {.}}{%
{\protect \APACyear {2019}}%
}]{%
duru2019}
\APACinsertmetastar {%
duru2019}%
\begin{APACrefauthors}%
{Duru}, F.%
, {Brain}, B.%
, {Gurnett}, D\BPBI A.%
, {Halekas}, J.%
, {Morgan}, D\BPBI D.%
\BCBL {}\ \BBA {} {Wilkinson}, C\BPBI J.%
\end{APACrefauthors}%
\unskip\
\newblock
\APACrefYearMonthDay{2019}{}{}.
\newblock
{\BBOQ}\APACrefatitle {{Electron Density Profiles in the Upper Ionosphere of
  {M}ars From 11 Years of {MARSIS} Data: {V}ariability Due to Seasons, Solar
  Cycle, and Crustal Magnetic Fields}} {{Electron Density Profiles in the Upper
  Ionosphere of {M}ars From 11 Years of {MARSIS} Data: {V}ariability Due to
  Seasons, Solar Cycle, and Crustal Magnetic Fields}}.{\BBCQ}
\newblock
\APACjournalVolNumPages{J. Geophys. Res.}{124}{4}{3057-3066}.
\newblock
\begin{APACrefDOI} \doi{10.1029/2018JA026327} \end{APACrefDOI}
\PrintBackRefs{\CurrentBib}

\bibitem [\protect \citeauthoryear {%
{Duru}%
\ \protect \BOthers {.}}{%
{Duru}%
\ \protect \BOthers {.}}{%
{\protect \APACyear {2009}}%
}]{%
duru2009}
\APACinsertmetastar {%
duru2009}%
\begin{APACrefauthors}%
{Duru}, F.%
, {Gurnett}, D\BPBI A.%
, {Frahm}, R\BPBI A.%
, {Winningham}, J\BPBI D.%
, {Morgan}, D\BPBI D.%
\BCBL {}\ \BBA {} {Howes}, G\BPBI G.%
\end{APACrefauthors}%
\unskip\
\newblock
\APACrefYearMonthDay{2009}{}{}.
\newblock
{\BBOQ}\APACrefatitle {{Steep, transient density gradients in the {M}artian
  ionosphere similar to the ionopause at {V}enus}} {{Steep, transient density
  gradients in the {M}artian ionosphere similar to the ionopause at
  {V}enus}}.{\BBCQ}
\newblock
\APACjournalVolNumPages{J. Geophys. Res.}{114}{}{A12310}.
\newblock
\begin{APACrefDOI} \doi{10.1029/2009JA014711} \end{APACrefDOI}
\PrintBackRefs{\CurrentBib}

\bibitem [\protect \citeauthoryear {%
{Duru}%
\ \protect \BOthers {.}}{%
{Duru}%
\ \protect \BOthers {.}}{%
{\protect \APACyear {2017}}%
}]{%
duru2017b}
\APACinsertmetastar {%
duru2017b}%
\begin{APACrefauthors}%
{Duru}, F.%
, {Gurnett}, D\BPBI A.%
, {Morgan}, D\BPBI D.%
, {Halekas}, J.%
, {Frahm}, R\BPBI A.%
, {Lundin}, R.%
\BDBL {}{Mahaffy}, P\BPBI R.%
\end{APACrefauthors}%
\unskip\
\newblock
\APACrefYearMonthDay{2017}{}{}.
\newblock
{\BBOQ}\APACrefatitle {{Response of the {M}artian ionosphere to solar activity
  including {SEP}s and {ICME}s in a two-week period starting on 25 {F}ebruary
  2015}} {{Response of the {M}artian ionosphere to solar activity including
  {SEP}s and {ICME}s in a two-week period starting on 25 {F}ebruary
  2015}}.{\BBCQ}
\newblock
\APACjournalVolNumPages{Planet. Space Sci.}{145}{}{28-37}.
\newblock
\begin{APACrefDOI} \doi{10.1016/j.pss.2017.07.010} \end{APACrefDOI}
\PrintBackRefs{\CurrentBib}

\bibitem [\protect \citeauthoryear {%
{Duru}%
\ \protect \BOthers {.}}{%
{Duru}%
\ \protect \BOthers {.}}{%
{\protect \APACyear {2008}}%
}]{%
duru2008}
\APACinsertmetastar {%
duru2008}%
\begin{APACrefauthors}%
{Duru}, F.%
, {Gurnett}, D\BPBI A.%
, {Morgan}, D\BPBI D.%
, {Modolo}, R.%
, {Nagy}, A\BPBI F.%
\BCBL {}\ \BBA {} {Najib}, D.%
\end{APACrefauthors}%
\unskip\
\newblock
\APACrefYearMonthDay{2008}{}{}.
\newblock
{\BBOQ}\APACrefatitle {Electron densities in the upper ionosphere of Mars from
  the excitation of electron plasma oscillations} {Electron densities in the
  upper ionosphere of mars from the excitation of electron plasma
  oscillations}.{\BBCQ}
\newblock
\APACjournalVolNumPages{J. Geophys. Res.}{113}{}{A07302, 10.1029/2008JA013073}.
\newblock
\begin{APACrefDOI} \doi{10.1029/2008JA013073} \end{APACrefDOI}
\PrintBackRefs{\CurrentBib}

\bibitem [\protect \citeauthoryear {%
{Edberg}%
, {Auster}%
\BCBL {}\ \protect \BOthers {.}}{%
{Edberg}%
, {Auster}%
\BCBL {}\ \protect \BOthers {.}}{%
{\protect \APACyear {2009}}%
}]{%
edberg2009b}
\APACinsertmetastar {%
edberg2009b}%
\begin{APACrefauthors}%
{Edberg}, N\BPBI J\BPBI T.%
, {Auster}, U.%
, {Barabash}, S.%
, {B{\"o}{\ss}wetter}, A.%
, {Brain}, D\BPBI A.%
, {Burch}, J\BPBI L.%
\BDBL {}{Trotignon}, J\BPBI G.%
\end{APACrefauthors}%
\unskip\
\newblock
\APACrefYearMonthDay{2009}{}{}.
\newblock
{\BBOQ}\APACrefatitle {{Rosetta and Mars Express observations of the influence
  of high solar wind pressure on the Martian plasma environment}} {{Rosetta and
  Mars Express observations of the influence of high solar wind pressure on the
  Martian plasma environment}}.{\BBCQ}
\newblock
\APACjournalVolNumPages{Ann. Geophys.}{27}{}{4533-4545}.
\newblock
\begin{APACrefDOI} \doi{10.5194/angeo-27-4533-2009} \end{APACrefDOI}
\PrintBackRefs{\CurrentBib}

\bibitem [\protect \citeauthoryear {%
{Edberg}%
, {Brain}%
\BCBL {}\ \protect \BOthers {.}}{%
{Edberg}%
, {Brain}%
\BCBL {}\ \protect \BOthers {.}}{%
{\protect \APACyear {2009}}%
}]{%
edberg2009}
\APACinsertmetastar {%
edberg2009}%
\begin{APACrefauthors}%
{Edberg}, N\BPBI J\BPBI T.%
, {Brain}, D\BPBI A.%
, {Lester}, M.%
, {Cowley}, S\BPBI W\BPBI H.%
, {Modolo}, R.%
, {Fr{\"a}nz}, M.%
\BCBL {}\ \BBA {} {Barabash}, S.%
\end{APACrefauthors}%
\unskip\
\newblock
\APACrefYearMonthDay{2009}{}{}.
\newblock
{\BBOQ}\APACrefatitle {{Plasma boundary variability at {M}ars as observed by
  {M}ars {G}lobal {S}urveyor and {M}ars {E}xpress}} {{Plasma boundary
  variability at {M}ars as observed by {M}ars {G}lobal {S}urveyor and {M}ars
  {E}xpress}}.{\BBCQ}
\newblock
\APACjournalVolNumPages{Annales Geophysicae}{27}{}{3537-3550}.
\newblock
\begin{APACrefDOI} \doi{10.5194/angeo-27-3537-2009} \end{APACrefDOI}
\PrintBackRefs{\CurrentBib}

\bibitem [\protect \citeauthoryear {%
{Edberg}%
\ \protect \BOthers {.}}{%
{Edberg}%
\ \protect \BOthers {.}}{%
{\protect \APACyear {2010}}%
}]{%
edberg2010}
\APACinsertmetastar {%
edberg2010}%
\begin{APACrefauthors}%
{Edberg}, N\BPBI J\BPBI T.%
, {Nilsson}, H.%
, {Williams}, A\BPBI O.%
, {Lester}, M.%
, {Milan}, S\BPBI E.%
, {Cowley}, S\BPBI W\BPBI H.%
\BDBL {}{Futaana}, Y.%
\end{APACrefauthors}%
\unskip\
\newblock
\APACrefYearMonthDay{2010}{}{}.
\newblock
{\BBOQ}\APACrefatitle {{Pumping out the atmosphere of Mars through solar wind
  pressure pulses}} {{Pumping out the atmosphere of Mars through solar wind
  pressure pulses}}.{\BBCQ}
\newblock
\APACjournalVolNumPages{Geophys. Res. Lett.}{37}{}{L03107}.
\newblock
\begin{APACrefDOI} \doi{10.1029/2009GL041814} \end{APACrefDOI}
\PrintBackRefs{\CurrentBib}

\bibitem [\protect \citeauthoryear {%
Ferguson%
, Cain%
, Crider%
, Brain%
\BCBL {}\ \BBA {} Harnett%
}{%
Ferguson%
\ \protect \BOthers {.}}{%
{\protect \APACyear {2005}}%
}]{%
ferguson2005}
\APACinsertmetastar {%
ferguson2005}%
\begin{APACrefauthors}%
Ferguson, B\BPBI B.%
, Cain, J\BPBI C.%
, Crider, D\BPBI H.%
, Brain, D\BPBI A.%
\BCBL {}\ \BBA {} Harnett, E\BPBI M.%
\end{APACrefauthors}%
\unskip\
\newblock
\APACrefYearMonthDay{2005}{}{}.
\newblock
{\BBOQ}\APACrefatitle {External fields on the nightside of Mars at {M}ars
  {G}lobal {S}urveyor mapping altitudes} {External fields on the nightside of
  mars at {M}ars {G}lobal {S}urveyor mapping altitudes}.{\BBCQ}
\newblock
\APACjournalVolNumPages{Geophys. Res. Lett.}{32}{16}{}.
\newblock
\begin{APACrefURL}
  \url{https://agupubs.onlinelibrary.wiley.com/doi/abs/10.1029/2004GL021964}
  \end{APACrefURL}
\newblock
\begin{APACrefDOI} \doi{10.1029/2004GL021964} \end{APACrefDOI}
\PrintBackRefs{\CurrentBib}

\bibitem [\protect \citeauthoryear {%
{Flynn}%
\ \protect \BOthers {.}}{%
{Flynn}%
\ \protect \BOthers {.}}{%
{\protect \APACyear {2017}}%
}]{%
flynn2017}
\APACinsertmetastar {%
flynn2017}%
\begin{APACrefauthors}%
{Flynn}, C\BPBI L.%
, {Vogt}, M\BPBI F.%
, {Withers}, P.%
, {Andersson}, L.%
, {England}, S.%
\BCBL {}\ \BBA {} {Liu}, G.%
\end{APACrefauthors}%
\unskip\
\newblock
\APACrefYearMonthDay{2017}{}{}.
\newblock
{\BBOQ}\APACrefatitle {{MAVEN Observations of the Effects of Crustal Magnetic
  Fields on Electron Density and Temperature in the Martian Dayside
  Ionosphere}} {{MAVEN Observations of the Effects of Crustal Magnetic Fields
  on Electron Density and Temperature in the Martian Dayside
  Ionosphere}}.{\BBCQ}
\newblock
\APACjournalVolNumPages{Geophys. Res. Lett.}{44}{21}{10,812-10,821}.
\newblock
\begin{APACrefDOI} \doi{10.1002/2017GL075367} \end{APACrefDOI}
\PrintBackRefs{\CurrentBib}

\bibitem [\protect \citeauthoryear {%
{Fowler}%
\ \protect \BOthers {.}}{%
{Fowler}%
\ \protect \BOthers {.}}{%
{\protect \APACyear {2018}}%
}]{%
fowler2018}
\APACinsertmetastar {%
fowler2018}%
\begin{APACrefauthors}%
{Fowler}, C\BPBI M.%
, {Andersson}, L.%
, {Ergun}, R\BPBI E.%
, {Harada}, Y.%
, {Hara}, T.%
, {Collinson}, G.%
\BDBL {}{Jakosky}, B\BPBI M.%
\end{APACrefauthors}%
\unskip\
\newblock
\APACrefYearMonthDay{2018}{}{}.
\newblock
{\BBOQ}\APACrefatitle {{{MAVEN} {O}bservations of {S}olar {W}ind-{D}riven
  {M}agnetosonic {W}aves {H}eating the {M}artian {D}ayside {I}onosphere}}
  {{{MAVEN} {O}bservations of {S}olar {W}ind-{D}riven {M}agnetosonic {W}aves
  {H}eating the {M}artian {D}ayside {I}onosphere}}.{\BBCQ}
\newblock
\APACjournalVolNumPages{J. Geophys. Res.}{123}{}{4129-4149}.
\newblock
\begin{APACrefDOI} \doi{10.1029/2018JA025208} \end{APACrefDOI}
\PrintBackRefs{\CurrentBib}

\bibitem [\protect \citeauthoryear {%
Fowler%
\ \protect \BOthers {.}}{%
Fowler%
\ \protect \BOthers {.}}{%
{\protect \APACyear {2019}}%
}]{%
fowler2019}
\APACinsertmetastar {%
fowler2019}%
\begin{APACrefauthors}%
Fowler, C\BPBI M.%
, Lee, C\BPBI O.%
, Xu, S.%
, Mitchell, D\BPBI L.%
, Lillis, R.%
, Weber, T.%
\BDBL {}Luhmann, J.%
\end{APACrefauthors}%
\unskip\
\newblock
\APACrefYearMonthDay{2019}{}{}.
\newblock
{\BBOQ}\APACrefatitle {The Penetration of Draped Magnetic Field Into the
  Martian Upper Ionosphere and Correlations With Upstream Solar Wind Dynamic
  Pressure} {The penetration of draped magnetic field into the martian upper
  ionosphere and correlations with upstream solar wind dynamic
  pressure}.{\BBCQ}
\newblock
\APACjournalVolNumPages{J. Geophys. Res., In Press}{0}{0}{}.
\newblock
\begin{APACrefDOI} \doi{10.1029/2019JA026550} \end{APACrefDOI}
\PrintBackRefs{\CurrentBib}

\bibitem [\protect \citeauthoryear {%
{Fox}%
}{%
{Fox}%
}{%
{\protect \APACyear {1997}}%
}]{%
fox1997}
\APACinsertmetastar {%
fox1997}%
\begin{APACrefauthors}%
{Fox}, J\BPBI L.%
\end{APACrefauthors}%
\unskip\
\newblock
\APACrefYearMonthDay{1997}{}{}.
\newblock
{\BBOQ}\APACrefatitle {{Upper limits to the outflow of ions at {M}ars:
  Implications for atmospheric evolution}} {{Upper limits to the outflow of
  ions at {M}ars: Implications for atmospheric evolution}}.{\BBCQ}
\newblock
\APACjournalVolNumPages{Geophys. Res. Lett.}{24}{}{2901}.
\newblock
\begin{APACrefDOI} \doi{10.1029/97GL52842} \end{APACrefDOI}
\PrintBackRefs{\CurrentBib}

\bibitem [\protect \citeauthoryear {%
{Fox}%
}{%
{Fox}%
}{%
{\protect \APACyear {2009}}%
}]{%
fox2009a}
\APACinsertmetastar {%
fox2009a}%
\begin{APACrefauthors}%
{Fox}, J\BPBI L.%
\end{APACrefauthors}%
\unskip\
\newblock
\APACrefYearMonthDay{2009}{}{}.
\newblock
{\BBOQ}\APACrefatitle {{Morphology of the dayside ionosphere of {M}ars:
  {I}mplications for ion outflows}} {{Morphology of the dayside ionosphere of
  {M}ars: {I}mplications for ion outflows}}.{\BBCQ}
\newblock
\APACjournalVolNumPages{J. Geophys. Res.}{114}{}{12005}.
\newblock
\begin{APACrefDOI} \doi{10.1029/2009JE003432} \end{APACrefDOI}
\PrintBackRefs{\CurrentBib}

\bibitem [\protect \citeauthoryear {%
{Fox}%
\ \BBA {} {Yeager}%
}{%
{Fox}%
\ \BBA {} {Yeager}%
}{%
{\protect \APACyear {2009}}%
}]{%
fox2009}
\APACinsertmetastar {%
fox2009}%
\begin{APACrefauthors}%
{Fox}, J\BPBI L.%
\BCBT {}\ \BBA {} {Yeager}, K\BPBI E.%
\end{APACrefauthors}%
\unskip\
\newblock
\APACrefYearMonthDay{2009}{{\APACmonth{04}}}{}.
\newblock
{\BBOQ}\APACrefatitle {{MGS} electron density profiles: Analysis of the peak
  magnitudes} {{MGS} electron density profiles: Analysis of the peak
  magnitudes}.{\BBCQ}
\newblock
\APACjournalVolNumPages{Icarus}{200}{}{468-479}.
\newblock
\begin{APACrefDOI} \doi{10.1016/j.icarus.2008.12.002} \end{APACrefDOI}
\PrintBackRefs{\CurrentBib}

\bibitem [\protect \citeauthoryear {%
Girazian%
, Mahaffy%
, Lee%
\BCBL {}\ \BBA {} Thiemann%
}{%
Girazian%
\ \protect \BOthers {.}}{%
{\protect \APACyear {2019}}%
}]{%
girazian2019}
\APACinsertmetastar {%
girazian2019}%
\begin{APACrefauthors}%
Girazian, Z.%
, Mahaffy, P.%
, Lee, Y.%
\BCBL {}\ \BBA {} Thiemann, E\BPBI M\BPBI B.%
\end{APACrefauthors}%
\unskip\
\newblock
\APACrefYearMonthDay{2019}{}{}.
\newblock
{\BBOQ}\APACrefatitle {Seasonal, Solar Zenith Angle, and Solar Flux Variations
  of {O$^+$} in the Topside Ionosphere of {M}ars} {Seasonal, solar zenith
  angle, and solar flux variations of {O$^+$} in the topside ionosphere of
  {M}ars}.{\BBCQ}
\newblock
\APACjournalVolNumPages{J. Geophys. Res., In Press}{0}{}{}.
\newblock
\begin{APACrefURL}
  \url{https://agupubs.onlinelibrary.wiley.com/doi/abs/10.1029/2018JA026086}
  \end{APACrefURL}
\newblock
\begin{APACrefDOI} \doi{10.1029/2018JA026086} \end{APACrefDOI}
\PrintBackRefs{\CurrentBib}

\bibitem [\protect \citeauthoryear {%
{Girazian}%
\ \protect \BOthers {.}}{%
{Girazian}%
\ \protect \BOthers {.}}{%
{\protect \APACyear {2017}}%
}]{%
girazian2017a}
\APACinsertmetastar {%
girazian2017a}%
\begin{APACrefauthors}%
{Girazian}, Z.%
, {Mahaffy}, P.%
, {Lillis}, R\BPBI J.%
, {Benna}, M.%
, {Elrod}, M.%
, {Fowler}, C\BPBI M.%
\BCBL {}\ \BBA {} {Mitchell}, D\BPBI L.%
\end{APACrefauthors}%
\unskip\
\newblock
\APACrefYearMonthDay{2017}{}{}.
\newblock
{\BBOQ}\APACrefatitle {{Ion {D}ensities in the {N}ightside {I}onosphere of
  {M}ars: {E}ffects of {E}lectron {I}mpact {I}onization}} {{Ion {D}ensities in
  the {N}ightside {I}onosphere of {M}ars: {E}ffects of {E}lectron {I}mpact
  {I}onization}}.{\BBCQ}
\newblock
\APACjournalVolNumPages{Geophys. Res. Lett.}{44}{}{11}.
\newblock
\begin{APACrefDOI} \doi{10.1002/2017GL075431} \end{APACrefDOI}
\PrintBackRefs{\CurrentBib}

\bibitem [\protect \citeauthoryear {%
{Girazian}%
\ \protect \BOthers {.}}{%
{Girazian}%
\ \protect \BOthers {.}}{%
{\protect \APACyear {2017}}%
}]{%
girazian2017}
\APACinsertmetastar {%
girazian2017}%
\begin{APACrefauthors}%
{Girazian}, Z.%
, {Mahaffy}, P\BPBI R.%
, {Lillis}, R\BPBI J.%
, {Benna}, M.%
, {Elrod}, M.%
\BCBL {}\ \BBA {} {Jakosky}, B\BPBI M.%
\end{APACrefauthors}%
\unskip\
\newblock
\APACrefYearMonthDay{2017}{}{}.
\newblock
{\BBOQ}\APACrefatitle {{Nightside ionosphere of {M}ars: {C}omposition, vertical
  structure, and variability}} {{Nightside ionosphere of {M}ars: {C}omposition,
  vertical structure, and variability}}.{\BBCQ}
\newblock
\APACjournalVolNumPages{J. Geophys. Res.}{122}{}{4712-4725}.
\newblock
\begin{APACrefDOI} \doi{10.1002/2016JA023508} \end{APACrefDOI}
\PrintBackRefs{\CurrentBib}

\bibitem [\protect \citeauthoryear {%
{Gurnett}%
\ \protect \BOthers {.}}{%
{Gurnett}%
\ \protect \BOthers {.}}{%
{\protect \APACyear {2008}}%
}]{%
gurnett2008}
\APACinsertmetastar {%
gurnett2008}%
\begin{APACrefauthors}%
{Gurnett}, D\BPBI A.%
, {Huff}, R\BPBI L.%
, {Morgan}, D\BPBI D.%
, {Persoon}, A\BPBI M.%
, {Averkamp}, T\BPBI F.%
, {Kirchner}, D\BPBI L.%
\BDBL {}{Picardi}, G.%
\end{APACrefauthors}%
\unskip\
\newblock
\APACrefYearMonthDay{2008}{}{}.
\newblock
{\BBOQ}\APACrefatitle {An overview of radar soundings of the martian ionosphere
  from the {M}ars {E}xpress spacecraft} {An overview of radar soundings of the
  martian ionosphere from the {M}ars {E}xpress spacecraft}.{\BBCQ}
\newblock
\APACjournalVolNumPages{Adv. Space Res.}{41}{}{1335-1346}.
\newblock
\begin{APACrefDOI} \doi{10.1016/j.asr.2007.01.062} \end{APACrefDOI}
\PrintBackRefs{\CurrentBib}

\bibitem [\protect \citeauthoryear {%
{Gurnett}%
\ \protect \BOthers {.}}{%
{Gurnett}%
\ \protect \BOthers {.}}{%
{\protect \APACyear {2005}}%
}]{%
gurnett2005}
\APACinsertmetastar {%
gurnett2005}%
\begin{APACrefauthors}%
{Gurnett}, D\BPBI A.%
, {Kirchner}, D\BPBI L.%
, {Huff}, R\BPBI L.%
, {Morgan}, D\BPBI D.%
, {Persoon}, A\BPBI M.%
, {Averkamp}, T\BPBI F.%
\BDBL {}{Picardi}, G.%
\end{APACrefauthors}%
\unskip\
\newblock
\APACrefYearMonthDay{2005}{}{}.
\newblock
{\BBOQ}\APACrefatitle {Radar soundings of the ionosphere of {M}ars} {Radar
  soundings of the ionosphere of {M}ars}.{\BBCQ}
\newblock
\APACjournalVolNumPages{Science}{310}{}{1929-1933}.
\newblock
\begin{APACrefDOI} \doi{10.1126/science.1121868} \end{APACrefDOI}
\PrintBackRefs{\CurrentBib}

\bibitem [\protect \citeauthoryear {%
{Halekas}%
, {Brain}%
\BCBL {}\ \protect \BOthers {.}}{%
{Halekas}%
, {Brain}%
\BCBL {}\ \protect \BOthers {.}}{%
{\protect \APACyear {2017}}%
}]{%
halekas2017a}
\APACinsertmetastar {%
halekas2017a}%
\begin{APACrefauthors}%
{Halekas}, J\BPBI S.%
, {Brain}, D\BPBI A.%
, {Luhmann}, J\BPBI G.%
, {DiBraccio}, G\BPBI A.%
, {Ruhunusiri}, S.%
, {Harada}, Y.%
\BDBL {}{Jakosky}, B\BPBI M.%
\end{APACrefauthors}%
\unskip\
\newblock
\APACrefYearMonthDay{2017}{}{}.
\newblock
{\BBOQ}\APACrefatitle {{Flows, Fields, and Forces in the Mars-Solar Wind
  Interaction}} {{Flows, Fields, and Forces in the Mars-Solar Wind
  Interaction}}.{\BBCQ}
\newblock
\APACjournalVolNumPages{J. Geophys. Res.}{122}{11}{11,320-11,341}.
\newblock
\begin{APACrefDOI} \doi{10.1002/2017JA024772} \end{APACrefDOI}
\PrintBackRefs{\CurrentBib}

\bibitem [\protect \citeauthoryear {%
{Halekas}%
\ \protect \BOthers {.}}{%
{Halekas}%
\ \protect \BOthers {.}}{%
{\protect \APACyear {2016}}%
}]{%
halekas2016}
\APACinsertmetastar {%
halekas2016}%
\begin{APACrefauthors}%
{Halekas}, J\BPBI S.%
, {Brain}, D\BPBI A.%
, {Ruhunusiri}, S.%
, {McFadden}, J\BPBI P.%
, {Mitchell}, D\BPBI L.%
, {Mazelle}, C.%
\BDBL {}{Jakosky}, B\BPBI M.%
\end{APACrefauthors}%
\unskip\
\newblock
\APACrefYearMonthDay{2016}{}{}.
\newblock
{\BBOQ}\APACrefatitle {{Plasma clouds and snowplows: {B}ulk plasma escape from
  Mars observed by {MAVEN}}} {{Plasma clouds and snowplows: {B}ulk plasma
  escape from Mars observed by {MAVEN}}}.{\BBCQ}
\newblock
\APACjournalVolNumPages{Geophys. Res. Lett.}{43}{4}{1426-1434}.
\newblock
\begin{APACrefDOI} \doi{10.1002/2016GL067752} \end{APACrefDOI}
\PrintBackRefs{\CurrentBib}

\bibitem [\protect \citeauthoryear {%
{Halekas}%
\ \protect \BOthers {.}}{%
{Halekas}%
\ \protect \BOthers {.}}{%
{\protect \APACyear {2018}}%
}]{%
halekas2018}
\APACinsertmetastar {%
halekas2018}%
\begin{APACrefauthors}%
{Halekas}, J\BPBI S.%
, {McFadden}, J\BPBI P.%
, {Brain}, D\BPBI A.%
, {Luhmann}, J\BPBI G.%
, {DiBraccio}, G\BPBI A.%
, {Connerney}, J\BPBI E\BPBI P.%
\BDBL {}{Jakosky}, B\BPBI M.%
\end{APACrefauthors}%
\unskip\
\newblock
\APACrefYearMonthDay{2018}{}{}.
\newblock
{\BBOQ}\APACrefatitle {{Structure and Variability of the {M}artian Ion
  Composition Boundary Layer}} {{Structure and Variability of the {M}artian Ion
  Composition Boundary Layer}}.{\BBCQ}
\newblock
\APACjournalVolNumPages{J. Geophys. Res.}{123}{10}{8439-8458}.
\newblock
\begin{APACrefDOI} \doi{10.1029/2018JA025866} \end{APACrefDOI}
\PrintBackRefs{\CurrentBib}

\bibitem [\protect \citeauthoryear {%
{Halekas}%
, {Ruhunusiri}%
\BCBL {}\ \protect \BOthers {.}}{%
{Halekas}%
, {Ruhunusiri}%
\BCBL {}\ \protect \BOthers {.}}{%
{\protect \APACyear {2017}}%
}]{%
halekas2017}
\APACinsertmetastar {%
halekas2017}%
\begin{APACrefauthors}%
{Halekas}, J\BPBI S.%
, {Ruhunusiri}, S.%
, {Harada}, Y.%
, {Collinson}, G.%
, {Mitchell}, D\BPBI L.%
, {Mazelle}, C.%
\BDBL {}{Jakosky}, B\BPBI M.%
\end{APACrefauthors}%
\unskip\
\newblock
\APACrefYearMonthDay{2017}{}{}.
\newblock
{\BBOQ}\APACrefatitle {{Structure, dynamics, and seasonal variability of the
  {M}ars-solar wind interaction: {MAVEN} {S}olar {W}ind {I}on {A}nalyzer
  in-flight performance and science results}} {{Structure, dynamics, and
  seasonal variability of the {M}ars-solar wind interaction: {MAVEN} {S}olar
  {W}ind {I}on {A}nalyzer in-flight performance and science results}}.{\BBCQ}
\newblock
\APACjournalVolNumPages{J. Geophys. Res.}{122}{}{547-578}.
\newblock
\begin{APACrefDOI} \doi{10.1002/2016JA023167} \end{APACrefDOI}
\PrintBackRefs{\CurrentBib}

\bibitem [\protect \citeauthoryear {%
{Halekas}%
\ \protect \BOthers {.}}{%
{Halekas}%
\ \protect \BOthers {.}}{%
{\protect \APACyear {2015}}%
}]{%
halekas2015a}
\APACinsertmetastar {%
halekas2015a}%
\begin{APACrefauthors}%
{Halekas}, J\BPBI S.%
, {Taylor}, E\BPBI R.%
, {Dalton}, G.%
, {Johnson}, G.%
, {Curtis}, D\BPBI W.%
, {McFadden}, J\BPBI P.%
\BDBL {}{Jakosky}, B\BPBI M.%
\end{APACrefauthors}%
\unskip\
\newblock
\APACrefYearMonthDay{2015}{}{}.
\newblock
{\BBOQ}\APACrefatitle {{The Solar Wind Ion Analyzer for MAVEN}} {{The Solar
  Wind Ion Analyzer for MAVEN}}.{\BBCQ}
\newblock
\APACjournalVolNumPages{Space Sci. Rev.}{195}{}{125-151}.
\newblock
\begin{APACrefDOI} \doi{10.1007/s11214-013-0029-z} \end{APACrefDOI}
\PrintBackRefs{\CurrentBib}

\bibitem [\protect \citeauthoryear {%
{Hanson}%
, {Sanatani}%
\BCBL {}\ \BBA {} {Zuccaro}%
}{%
{Hanson}%
\ \protect \BOthers {.}}{%
{\protect \APACyear {1977}}%
}]{%
hanson1977}
\APACinsertmetastar {%
hanson1977}%
\begin{APACrefauthors}%
{Hanson}, W\BPBI B.%
, {Sanatani}, S.%
\BCBL {}\ \BBA {} {Zuccaro}, D\BPBI R.%
\end{APACrefauthors}%
\unskip\
\newblock
\APACrefYearMonthDay{1977}{}{}.
\newblock
{\BBOQ}\APACrefatitle {The martian ionosphere as observed by the {V}iking
  {R}etarding {P}otential {A}nalyzers} {The martian ionosphere as observed by
  the {V}iking {R}etarding {P}otential {A}nalyzers}.{\BBCQ}
\newblock
\APACjournalVolNumPages{J. Geophys. Res.}{82}{}{4351-4363}.
\newblock
\begin{APACrefDOI} \doi{10.1029/JS082i028p04351} \end{APACrefDOI}
\PrintBackRefs{\CurrentBib}

\bibitem [\protect \citeauthoryear {%
Harada%
\ \protect \BOthers {.}}{%
Harada%
\ \protect \BOthers {.}}{%
{\protect \APACyear {2018}}%
}]{%
harada2018}
\APACinsertmetastar {%
harada2018}%
\begin{APACrefauthors}%
Harada, Y.%
, Gurnett, D\BPBI A.%
, Kopf, A\BPBI J.%
, Halekas, J\BPBI S.%
, Ruhunusiri, S.%
, DiBraccio, G\BPBI A.%
\BDBL {}Brain, D\BPBI A.%
\end{APACrefauthors}%
\unskip\
\newblock
\APACrefYearMonthDay{2018}{}{}.
\newblock
{\BBOQ}\APACrefatitle {{MARSIS} Observations of the Martian Nightside
  Ionosphere During the {S}eptember 2017 Solar Event} {{MARSIS} observations of
  the martian nightside ionosphere during the {S}eptember 2017 solar
  event}.{\BBCQ}
\newblock
\APACjournalVolNumPages{Geophys. Res. Lett.}{45}{16}{7960-7967}.
\newblock
\begin{APACrefURL}
  \url{https://agupubs.onlinelibrary.wiley.com/doi/abs/10.1002/2018GL077622}
  \end{APACrefURL}
\newblock
\begin{APACrefDOI} \doi{10.1002/2018GL077622} \end{APACrefDOI}
\PrintBackRefs{\CurrentBib}

\bibitem [\protect \citeauthoryear {%
{Harada}%
\ \protect \BOthers {.}}{%
{Harada}%
\ \protect \BOthers {.}}{%
{\protect \APACyear {2017}}%
}]{%
harada2017}
\APACinsertmetastar {%
harada2017}%
\begin{APACrefauthors}%
{Harada}, Y.%
, {Gurnett}, D\BPBI A.%
, {Kopf}, A\BPBI J.%
, {Halekas}, J\BPBI S.%
, {Ruhunusiri}, S.%
, {Lee}, C\BPBI O.%
\BDBL {}{Jakosky}, B\BPBI M.%
\end{APACrefauthors}%
\unskip\
\newblock
\APACrefYearMonthDay{2017}{}{}.
\newblock
{\BBOQ}\APACrefatitle {{Dynamic response of the Martian ionosphere to an
  interplanetary shock: {M}ars {E}xpress and {MAVEN} observations}} {{Dynamic
  response of the Martian ionosphere to an interplanetary shock: {M}ars
  {E}xpress and {MAVEN} observations}}.{\BBCQ}
\newblock
\APACjournalVolNumPages{Geophys. Res. Lett.}{44}{18}{9116-9123}.
\newblock
\begin{APACrefDOI} \doi{10.1002/2017GL074897} \end{APACrefDOI}
\PrintBackRefs{\CurrentBib}

\bibitem [\protect \citeauthoryear {%
{Jakosky}%
, {Grebowsky}%
\BCBL {}\ \protect \BOthers {.}}{%
{Jakosky}%
, {Grebowsky}%
\BCBL {}\ \protect \BOthers {.}}{%
{\protect \APACyear {2015}}%
}]{%
jakosky2015b}
\APACinsertmetastar {%
jakosky2015b}%
\begin{APACrefauthors}%
{Jakosky}, B\BPBI M.%
, {Grebowsky}, J\BPBI M.%
, {Luhmann}, J\BPBI G.%
, {Connerney}, J.%
, {Eparvier}, F.%
, {Ergun}, R.%
\BDBL {}{Yelle}, R.%
\end{APACrefauthors}%
\unskip\
\newblock
\APACrefYearMonthDay{2015}{}{}.
\newblock
{\BBOQ}\APACrefatitle {{{MAVEN} observations of the response of {M}ars to an
  interplanetary coronal mass ejection}} {{{MAVEN} observations of the response
  of {M}ars to an interplanetary coronal mass ejection}}.{\BBCQ}
\newblock
\APACjournalVolNumPages{Science}{350}{6261}{0210}.
\newblock
\begin{APACrefDOI} \doi{10.1126/science.aad0210} \end{APACrefDOI}
\PrintBackRefs{\CurrentBib}

\bibitem [\protect \citeauthoryear {%
{Jakosky}%
, {Lin}%
\BCBL {}\ \protect \BOthers {.}}{%
{Jakosky}%
, {Lin}%
\BCBL {}\ \protect \BOthers {.}}{%
{\protect \APACyear {2015}}%
}]{%
jakosky2015}
\APACinsertmetastar {%
jakosky2015}%
\begin{APACrefauthors}%
{Jakosky}, B\BPBI M.%
, {Lin}, R\BPBI P.%
, {Grebowsky}, J\BPBI M.%
, {Luhmann}, J\BPBI G.%
, {Mitchell}, D\BPBI F.%
, {Beutelschies}, G.%
\BDBL {}{Zurek}, R.%
\end{APACrefauthors}%
\unskip\
\newblock
\APACrefYearMonthDay{2015}{}{}.
\newblock
{\BBOQ}\APACrefatitle {{The Mars Atmosphere and Volatile Evolution (MAVEN)
  Mission}} {{The Mars Atmosphere and Volatile Evolution (MAVEN)
  Mission}}.{\BBCQ}
\newblock
\APACjournalVolNumPages{Space Sci. Rev.}{}{}{}.
\newblock
\begin{APACrefDOI} \doi{10.1007/s11214-015-0139-x} \end{APACrefDOI}
\PrintBackRefs{\CurrentBib}

\bibitem [\protect \citeauthoryear {%
{Kar}%
, {Mahajan}%
\BCBL {}\ \BBA {} {Kohli}%
}{%
{Kar}%
\ \protect \BOthers {.}}{%
{\protect \APACyear {1996}}%
}]{%
kar1996a}
\APACinsertmetastar {%
kar1996a}%
\begin{APACrefauthors}%
{Kar}, J.%
, {Mahajan}, K\BPBI K.%
\BCBL {}\ \BBA {} {Kohli}, R.%
\end{APACrefauthors}%
\unskip\
\newblock
\APACrefYearMonthDay{1996}{}{}.
\newblock
{\BBOQ}\APACrefatitle {{On the outflow of O$^{+}_{2}$ ions at {M}ars}} {{On the
  outflow of O$^{+}_{2}$ ions at {M}ars}}.{\BBCQ}
\newblock
\APACjournalVolNumPages{J. Geophys. Res.}{101}{}{12747-12752}.
\newblock
\begin{APACrefDOI} \doi{10.1029/95JE03526} \end{APACrefDOI}
\PrintBackRefs{\CurrentBib}

\bibitem [\protect \citeauthoryear {%
{Lee}%
\ \protect \BOthers {.}}{%
{Lee}%
\ \protect \BOthers {.}}{%
{\protect \APACyear {2017}}%
}]{%
lee2017}
\APACinsertmetastar {%
lee2017}%
\begin{APACrefauthors}%
{Lee}, C\BPBI O.%
, {Hara}, T.%
, {Halekas}, J\BPBI S.%
, {Thiemann}, E.%
, {Chamberlin}, P.%
, {Eparvier}, F.%
\BDBL {}{Jakosky}, B\BPBI M.%
\end{APACrefauthors}%
\unskip\
\newblock
\APACrefYearMonthDay{2017}{}{}.
\newblock
{\BBOQ}\APACrefatitle {{{MAVEN} observations of the solar cycle 24 space
  weather conditions at {M}ars}} {{{MAVEN} observations of the solar cycle 24
  space weather conditions at {M}ars}}.{\BBCQ}
\newblock
\APACjournalVolNumPages{J. Geophys. Res.}{122}{}{2768-2794}.
\newblock
\begin{APACrefDOI} \doi{10.1002/2016JA023495} \end{APACrefDOI}
\PrintBackRefs{\CurrentBib}

\bibitem [\protect \citeauthoryear {%
{Luhmann}%
\ \protect \BOthers {.}}{%
{Luhmann}%
\ \protect \BOthers {.}}{%
{\protect \APACyear {2017}}%
}]{%
luhmann2017}
\APACinsertmetastar {%
luhmann2017}%
\begin{APACrefauthors}%
{Luhmann}, J\BPBI G.%
, {Dong}, C\BPBI F.%
, {Ma}, Y\BPBI J.%
, {Curry}, S\BPBI M.%
, {Xu}, S.%
, {Lee}, C\BPBI O.%
\BDBL {}{Jakosky}, B\BPBI M.%
\end{APACrefauthors}%
\unskip\
\newblock
\APACrefYearMonthDay{2017}{}{}.
\newblock
{\BBOQ}\APACrefatitle {{Martian magnetic storms}} {{Martian magnetic
  storms}}.{\BBCQ}
\newblock
\APACjournalVolNumPages{J. geophys. Res.}{122}{6}{6185-6209}.
\newblock
\begin{APACrefDOI} \doi{10.1002/2016JA023513} \end{APACrefDOI}
\PrintBackRefs{\CurrentBib}

\bibitem [\protect \citeauthoryear {%
{Luhmann}%
, {Russell}%
, {Scarf}%
, {Brace}%
\BCBL {}\ \BBA {} {Knudsen}%
}{%
{Luhmann}%
\ \protect \BOthers {.}}{%
{\protect \APACyear {1987}}%
}]{%
luhmann1987}
\APACinsertmetastar {%
luhmann1987}%
\begin{APACrefauthors}%
{Luhmann}, J\BPBI G.%
, {Russell}, C\BPBI T.%
, {Scarf}, F\BPBI L.%
, {Brace}, L\BPBI H.%
\BCBL {}\ \BBA {} {Knudsen}, W\BPBI C.%
\end{APACrefauthors}%
\unskip\
\newblock
\APACrefYearMonthDay{1987}{}{}.
\newblock
{\BBOQ}\APACrefatitle {{Characteristics of the {M}arslike limit of the
  {V}enus-solar wind interaction}} {{Characteristics of the {M}arslike limit of
  the {V}enus-solar wind interaction}}.{\BBCQ}
\newblock
\APACjournalVolNumPages{J. Geophys. Res.}{92}{}{8545-8557}.
\newblock
\begin{APACrefDOI} \doi{10.1029/JA092iA08p08545} \end{APACrefDOI}
\PrintBackRefs{\CurrentBib}

\bibitem [\protect \citeauthoryear {%
Ma%
, Nagy%
, Sokolov%
\BCBL {}\ \BBA {} Hansen%
}{%
Ma%
\ \protect \BOthers {.}}{%
{\protect \APACyear {2004}}%
}]{%
ma2004}
\APACinsertmetastar {%
ma2004}%
\begin{APACrefauthors}%
Ma, Y.%
, Nagy, A\BPBI F.%
, Sokolov, I\BPBI V.%
\BCBL {}\ \BBA {} Hansen, K\BPBI C.%
\end{APACrefauthors}%
\unskip\
\newblock
\APACrefYearMonthDay{2004}{}{}.
\newblock
{\BBOQ}\APACrefatitle {Three-dimensional, multispecies, high spatial resolution
  {MHD} studies of the solar wind interaction with {M}ars} {Three-dimensional,
  multispecies, high spatial resolution {MHD} studies of the solar wind
  interaction with {M}ars}.{\BBCQ}
\newblock
\APACjournalVolNumPages{J. Geophys. Res.}{109}{}{A07211, 10.1029/2003JA010367}.
\newblock
\begin{APACrefDOI} \doi{10.1029/2003JA010367} \end{APACrefDOI}
\PrintBackRefs{\CurrentBib}

\bibitem [\protect \citeauthoryear {%
{Ma}%
, {Fang}%
, {Nagy}%
, {Russell}%
\BCBL {}\ \BBA {} {Toth}%
}{%
{Ma}%
\ \protect \BOthers {.}}{%
{\protect \APACyear {2014}}%
}]{%
ma2014}
\APACinsertmetastar {%
ma2014}%
\begin{APACrefauthors}%
{Ma}, Y\BPBI J.%
, {Fang}, X.%
, {Nagy}, A\BPBI F.%
, {Russell}, C\BPBI T.%
\BCBL {}\ \BBA {} {Toth}, G.%
\end{APACrefauthors}%
\unskip\
\newblock
\APACrefYearMonthDay{2014}{}{}.
\newblock
{\BBOQ}\APACrefatitle {{Martian ionospheric responses to dynamic pressure
  enhancements in the solar wind}} {{Martian ionospheric responses to dynamic
  pressure enhancements in the solar wind}}.{\BBCQ}
\newblock
\APACjournalVolNumPages{J. Geophys. Res.}{119}{2}{1272-1286}.
\newblock
\begin{APACrefDOI} \doi{10.1002/2013JA019402} \end{APACrefDOI}
\PrintBackRefs{\CurrentBib}

\bibitem [\protect \citeauthoryear {%
{Ma}%
\ \protect \BOthers {.}}{%
{Ma}%
\ \protect \BOthers {.}}{%
{\protect \APACyear {2017}}%
}]{%
ma2017}
\APACinsertmetastar {%
ma2017}%
\begin{APACrefauthors}%
{Ma}, Y\BPBI J.%
, {Russell}, C\BPBI T.%
, {Fang}, X.%
, {Dong}, C\BPBI F.%
, {Nagy}, A\BPBI F.%
, {Toth}, G.%
\BDBL {}{Jakosky}, B\BPBI M.%
\end{APACrefauthors}%
\unskip\
\newblock
\APACrefYearMonthDay{2017}{}{}.
\newblock
{\BBOQ}\APACrefatitle {{Variations of the {M}artian plasma environment during
  the {ICME} passage on 8 {M}arch 2015: {A} time-dependent {MHD} study}}
  {{Variations of the {M}artian plasma environment during the {ICME} passage on
  8 {M}arch 2015: {A} time-dependent {MHD} study}}.{\BBCQ}
\newblock
\APACjournalVolNumPages{J. Geophys. Res.}{122}{2}{1714-1730}.
\newblock
\begin{APACrefDOI} \doi{10.1002/2016JA023402} \end{APACrefDOI}
\PrintBackRefs{\CurrentBib}

\bibitem [\protect \citeauthoryear {%
{Mendillo}%
\ \protect \BOthers {.}}{%
{Mendillo}%
\ \protect \BOthers {.}}{%
{\protect \APACyear {2017}}%
}]{%
mendillo2017}
\APACinsertmetastar {%
mendillo2017}%
\begin{APACrefauthors}%
{Mendillo}, M.%
, {Narvaez}, C.%
, {Vogt}, M\BPBI F.%
, {Mayyasi}, M.%
, {Mahaffy}, P.%
, {Benna}, M.%
\BDBL {}{Jakosky}, B.%
\end{APACrefauthors}%
\unskip\
\newblock
\APACrefYearMonthDay{2017}{}{}.
\newblock
{\BBOQ}\APACrefatitle {{{MAVEN} and the total electron content of the {M}artian
  ionosphere}} {{{MAVEN} and the total electron content of the {M}artian
  ionosphere}}.{\BBCQ}
\newblock
\APACjournalVolNumPages{J. Geophys. Res.}{122}{3}{3526-3537}.
\newblock
\begin{APACrefDOI} \doi{10.1002/2016JA023474} \end{APACrefDOI}
\PrintBackRefs{\CurrentBib}

\bibitem [\protect \citeauthoryear {%
{Miller}%
, {Knudsen}%
\BCBL {}\ \BBA {} {Spenner}%
}{%
{Miller}%
\ \protect \BOthers {.}}{%
{\protect \APACyear {1984}}%
}]{%
miller1984}
\APACinsertmetastar {%
miller1984}%
\begin{APACrefauthors}%
{Miller}, K\BPBI L.%
, {Knudsen}, W\BPBI C.%
\BCBL {}\ \BBA {} {Spenner}, K.%
\end{APACrefauthors}%
\unskip\
\newblock
\APACrefYearMonthDay{1984}{{\APACmonth{03}}}{}.
\newblock
{\BBOQ}\APACrefatitle {The dayside Venus ionosphere. I - Pioneer-Venus
  retarding potential analyzer experimental observations} {The dayside venus
  ionosphere. i - pioneer-venus retarding potential analyzer experimental
  observations}.{\BBCQ}
\newblock
\APACjournalVolNumPages{Icarus}{57}{}{386-409}.
\newblock
\begin{APACrefDOI} \doi{10.1016/0019-1035(84)90125-8} \end{APACrefDOI}
\PrintBackRefs{\CurrentBib}

\bibitem [\protect \citeauthoryear {%
{Morgan}%
\ \protect \BOthers {.}}{%
{Morgan}%
\ \protect \BOthers {.}}{%
{\protect \APACyear {2014}}%
}]{%
morgan2014}
\APACinsertmetastar {%
morgan2014}%
\begin{APACrefauthors}%
{Morgan}, D\BPBI D.%
, {Di{\'e}val}, C.%
, {Gurnett}, D\BPBI A.%
, {Duru}, F.%
, {Dubinin}, E\BPBI M.%
, {Fr{\"a}nz}, M.%
\BDBL {}{Plaut}, J\BPBI J.%
\end{APACrefauthors}%
\unskip\
\newblock
\APACrefYearMonthDay{2014}{Jul}{}.
\newblock
{\BBOQ}\APACrefatitle {{Effects of a strong {ICME} on the Martian ionosphere as
  detected by {M}ars {E}xpress and {M}ars {O}dyssey}} {{Effects of a strong
  {ICME} on the Martian ionosphere as detected by {M}ars {E}xpress and {M}ars
  {O}dyssey}}.{\BBCQ}
\newblock
\APACjournalVolNumPages{J. Geophys. Res.}{119}{7}{5891-5908}.
\newblock
\begin{APACrefDOI} \doi{10.1002/2013JA019522} \end{APACrefDOI}
\PrintBackRefs{\CurrentBib}

\bibitem [\protect \citeauthoryear {%
{Morschhauser}%
, {Lesur}%
\BCBL {}\ \BBA {} {Grott}%
}{%
{Morschhauser}%
\ \protect \BOthers {.}}{%
{\protect \APACyear {2014}}%
}]{%
morsch2014}
\APACinsertmetastar {%
morsch2014}%
\begin{APACrefauthors}%
{Morschhauser}, A.%
, {Lesur}, V.%
\BCBL {}\ \BBA {} {Grott}, M.%
\end{APACrefauthors}%
\unskip\
\newblock
\APACrefYearMonthDay{2014}{}{}.
\newblock
{\BBOQ}\APACrefatitle {{A spherical harmonic model of the lithospheric magnetic
  field of {M}ars}} {{A spherical harmonic model of the lithospheric magnetic
  field of {M}ars}}.{\BBCQ}
\newblock
\APACjournalVolNumPages{J. Geophys. Res.}{119}{}{1162-1188}.
\newblock
\begin{APACrefDOI} \doi{10.1002/2013JE004555} \end{APACrefDOI}
\PrintBackRefs{\CurrentBib}

\bibitem [\protect \citeauthoryear {%
{Nagy}%
\ \protect \BOthers {.}}{%
{Nagy}%
\ \protect \BOthers {.}}{%
{\protect \APACyear {2004}}%
}]{%
nagy2004}
\APACinsertmetastar {%
nagy2004}%
\begin{APACrefauthors}%
{Nagy}, A\BPBI F.%
, {Winterhalter}, D.%
, {Sauer}, K.%
, {Cravens}, T\BPBI E.%
, {Brecht}, S.%
, {Mazelle}, C.%
\BDBL {}{Trotignon}, J\BPBI G.%
\end{APACrefauthors}%
\unskip\
\newblock
\APACrefYearMonthDay{2004}{}{}.
\newblock
{\BBOQ}\APACrefatitle {{The plasma Environment of Mars}} {{The plasma
  Environment of Mars}}.{\BBCQ}
\newblock
\APACjournalVolNumPages{Space Sci. Rev.}{111}{}{33-114}.
\newblock
\begin{APACrefDOI} \doi{10.1023/B:SPAC.0000032718.47512.92} \end{APACrefDOI}
\PrintBackRefs{\CurrentBib}

\bibitem [\protect \citeauthoryear {%
{N{\v{e}}mec}%
, {Morgan}%
, {Gurnett}%
\BCBL {}\ \BBA {} {Andrews}%
}{%
{N{\v{e}}mec}%
\ \protect \BOthers {.}}{%
{\protect \APACyear {2016}}%
}]{%
nemec2016}
\APACinsertmetastar {%
nemec2016}%
\begin{APACrefauthors}%
{N{\v{e}}mec}, F.%
, {Morgan}, D\BPBI D.%
, {Gurnett}, D\BPBI A.%
\BCBL {}\ \BBA {} {Andrews}, D\BPBI J.%
\end{APACrefauthors}%
\unskip\
\newblock
\APACrefYearMonthDay{2016}{}{}.
\newblock
{\BBOQ}\APACrefatitle {{Empirical model of the Martian dayside ionosphere:
  Effects of crustal magnetic fields and solar ionizing flux at higher
  altitudes}} {{Empirical model of the Martian dayside ionosphere: Effects of
  crustal magnetic fields and solar ionizing flux at higher altitudes}}.{\BBCQ}
\newblock
\APACjournalVolNumPages{J. Geophys. Res.}{121}{2}{1760-1771}.
\newblock
\begin{APACrefDOI} \doi{10.1002/2015JA022060} \end{APACrefDOI}
\PrintBackRefs{\CurrentBib}

\bibitem [\protect \citeauthoryear {%
{N{\v{e}}mec}%
\ \protect \BOthers {.}}{%
{N{\v{e}}mec}%
\ \protect \BOthers {.}}{%
{\protect \APACyear {2019}}%
}]{%
nemec2019}
\APACinsertmetastar {%
nemec2019}%
\begin{APACrefauthors}%
{N{\v{e}}mec}, F.%
, {Morgan}, D\BPBI D.%
, {Kopf}, A\BPBI J.%
, {Gurnett}, D\BPBI A.%
, {Pito{\r{A}}{\'a}k}, D.%
, {Fowler}, C\BPBI M.%
\BDBL {}{Andersson}, L.%
\end{APACrefauthors}%
\unskip\
\newblock
\APACrefYearMonthDay{2019}{}{}.
\newblock
{\BBOQ}\APACrefatitle {{Characterizing Average Electron Densities in the
  Martian Dayside Upper Ionosphere}} {{Characterizing Average Electron
  Densities in the Martian Dayside Upper Ionosphere}}.{\BBCQ}
\newblock
\APACjournalVolNumPages{J. Geophys. Res.}{124}{1}{76-93}.
\newblock
\begin{APACrefDOI} \doi{10.1029/2018JE005849} \end{APACrefDOI}
\PrintBackRefs{\CurrentBib}

\bibitem [\protect \citeauthoryear {%
{Opgenoorth}%
\ \protect \BOthers {.}}{%
{Opgenoorth}%
\ \protect \BOthers {.}}{%
{\protect \APACyear {2013}}%
}]{%
opgen2013}
\APACinsertmetastar {%
opgen2013}%
\begin{APACrefauthors}%
{Opgenoorth}, H\BPBI J.%
, {Andrews}, D\BPBI J.%
, {Fr{\"a}nz}, M.%
, {Lester}, M.%
, {Edberg}, N\BPBI J\BPBI T.%
, {Morgan}, D.%
\BDBL {}{Williams}, A\BPBI O.%
\end{APACrefauthors}%
\unskip\
\newblock
\APACrefYearMonthDay{2013}{}{}.
\newblock
{\BBOQ}\APACrefatitle {{Mars ionospheric response to solar wind variability}}
  {{Mars ionospheric response to solar wind variability}}.{\BBCQ}
\newblock
\APACjournalVolNumPages{J. Geophys. Res.}{118}{10}{6558-6587}.
\newblock
\begin{APACrefDOI} \doi{10.1002/jgra.50537} \end{APACrefDOI}
\PrintBackRefs{\CurrentBib}

\bibitem [\protect \citeauthoryear {%
{P{\'e}rez-de-Tejada}%
, {Lundin}%
, {Durand-Manterola}%
\BCBL {}\ \BBA {} {Reyes-Ruiz}%
}{%
{P{\'e}rez-de-Tejada}%
\ \protect \BOthers {.}}{%
{\protect \APACyear {2009}}%
}]{%
perez2009}
\APACinsertmetastar {%
perez2009}%
\begin{APACrefauthors}%
{P{\'e}rez-de-Tejada}, H.%
, {Lundin}, R.%
, {Durand-Manterola}, H.%
\BCBL {}\ \BBA {} {Reyes-Ruiz}, M.%
\end{APACrefauthors}%
\unskip\
\newblock
\APACrefYearMonthDay{2009}{}{}.
\newblock
{\BBOQ}\APACrefatitle {{Solar wind erosion of the polar regions of the {M}ars
  ionosphere}} {{Solar wind erosion of the polar regions of the {M}ars
  ionosphere}}.{\BBCQ}
\newblock
\APACjournalVolNumPages{J. Geophys. Res.}{114}{}{A02106}.
\newblock
\begin{APACrefDOI} \doi{10.1029/2008JA013295} \end{APACrefDOI}
\PrintBackRefs{\CurrentBib}

\bibitem [\protect \citeauthoryear {%
{Picardi}%
\ \protect \BOthers {.}}{%
{Picardi}%
\ \protect \BOthers {.}}{%
{\protect \APACyear {2004}}%
}]{%
picardi2004}
\APACinsertmetastar {%
picardi2004}%
\begin{APACrefauthors}%
{Picardi}, G.%
, {Biccari}, D.%
, {Seu}, R.%
, {Plaut}, J.%
, {Johnson}, W\BPBI T\BPBI K.%
, {Jordan}, R\BPBI L.%
\BDBL {}{Zampolini}, E.%
\end{APACrefauthors}%
\unskip\
\newblock
\APACrefYearMonthDay{2004}{}{}.
\newblock
{\BBOQ}\APACrefatitle {MARSIS: Mars Advanced Radar for Subsurface and
  Ionosphere Sounding} {Marsis: Mars advanced radar for subsurface and
  ionosphere sounding}.{\BBCQ}
\newblock
\BIn{} (\BPG~51-69).
\newblock
\APACaddressPublisher{}{ESA SP-1240: Mars Express: the Scientific Payload,
  available online at
  http://sci.esa.int/science-e/www/object/index.cfm?fobjectid=34885}.
\PrintBackRefs{\CurrentBib}

\bibitem [\protect \citeauthoryear {%
{Ramstad}%
\ \protect \BOthers {.}}{%
{Ramstad}%
\ \protect \BOthers {.}}{%
{\protect \APACyear {2017}}%
}]{%
ramstad2017}
\APACinsertmetastar {%
ramstad2017}%
\begin{APACrefauthors}%
{Ramstad}, R.%
, {Barabash}, S.%
, {Futaana}, Y.%
, {Yamauchi}, M.%
, {Nilsson}, H.%
\BCBL {}\ \BBA {} {Holmstr{\"o}m}, M.%
\end{APACrefauthors}%
\unskip\
\newblock
\APACrefYearMonthDay{2017}{}{}.
\newblock
{\BBOQ}\APACrefatitle {{Mars Under Primordial Solar Wind Conditions: {M}ars
  {E}xpress Observations of the Strongest {CME} Detected at {M}ars Under
  {S}olar {C}ycle \#24 and its Impact on Atmospheric Ion Escape}} {{Mars Under
  Primordial Solar Wind Conditions: {M}ars {E}xpress Observations of the
  Strongest {CME} Detected at {M}ars Under {S}olar {C}ycle \#24 and its Impact
  on Atmospheric Ion Escape}}.{\BBCQ}
\newblock
\APACjournalVolNumPages{Geophys. Res. Lett.}{44}{21}{10,805-10,811}.
\newblock
\begin{APACrefDOI} \doi{10.1002/2017GL075446} \end{APACrefDOI}
\PrintBackRefs{\CurrentBib}

\bibitem [\protect \citeauthoryear {%
{S{\'a}nchez-Cano}%
\ \protect \BOthers {.}}{%
{S{\'a}nchez-Cano}%
\ \protect \BOthers {.}}{%
{\protect \APACyear {2017}}%
}]{%
sanchez2017}
\APACinsertmetastar {%
sanchez2017}%
\begin{APACrefauthors}%
{S{\'a}nchez-Cano}, B.%
, {Hall}, B\BPBI E\BPBI S.%
, {Lester}, M.%
, {Mays}, M\BPBI L.%
, {Witasse}, O.%
, {Ambrosi}, R.%
\BDBL {}{Reyes-Ayala}, K\BPBI I.%
\end{APACrefauthors}%
\unskip\
\newblock
\APACrefYearMonthDay{2017}{Jun}{}.
\newblock
{\BBOQ}\APACrefatitle {{Mars plasma system response to solar wind disturbances
  during solar minimum}} {{Mars plasma system response to solar wind
  disturbances during solar minimum}}.{\BBCQ}
\newblock
\APACjournalVolNumPages{J. Geophys. Res.}{122}{6}{6611-6634}.
\newblock
\begin{APACrefDOI} \doi{10.1002/2016JA023587} \end{APACrefDOI}
\PrintBackRefs{\CurrentBib}

\bibitem [\protect \citeauthoryear {%
{Thampi}%
\ \protect \BOthers {.}}{%
{Thampi}%
\ \protect \BOthers {.}}{%
{\protect \APACyear {2018}}%
}]{%
thampi2018}
\APACinsertmetastar {%
thampi2018}%
\begin{APACrefauthors}%
{Thampi}, S\BPBI V.%
, {Krishnaprasad}, C.%
, {Bhardwaj}, A.%
, {Lee}, Y.%
, {Choudhary}, R\BPBI K.%
\BCBL {}\ \BBA {} {Pant}, T\BPBI K.%
\end{APACrefauthors}%
\unskip\
\newblock
\APACrefYearMonthDay{2018}{}{}.
\newblock
{\BBOQ}\APACrefatitle {{{MAVEN} Observations of the Response of Martian
  Ionosphere to the Interplanetary Coronal Mass Ejections of March 2015}}
  {{{MAVEN} Observations of the Response of Martian Ionosphere to the
  Interplanetary Coronal Mass Ejections of March 2015}}.{\BBCQ}
\newblock
\APACjournalVolNumPages{J. Geophys. Res.}{123}{8}{6917-6929}.
\newblock
\begin{APACrefDOI} \doi{10.1029/2018JA025444} \end{APACrefDOI}
\PrintBackRefs{\CurrentBib}

\bibitem [\protect \citeauthoryear {%
{Withers}%
\ \protect \BOthers {.}}{%
{Withers}%
\ \protect \BOthers {.}}{%
{\protect \APACyear {2012}}%
}]{%
withers2012}
\APACinsertmetastar {%
withers2012}%
\begin{APACrefauthors}%
{Withers}, P.%
, {Fallows}, K.%
, {Girazian}, Z.%
, {Matta}, M.%
, {H{\"a}usler}, B.%
, {Hinson}, D.%
\BDBL {}{Witasse}, O.%
\end{APACrefauthors}%
\unskip\
\newblock
\APACrefYearMonthDay{2012}{{\APACmonth{09}}}{}.
\newblock
{\BBOQ}\APACrefatitle {A clear view of the multifaceted dayside ionosphere of
  {M}ars} {A clear view of the multifaceted dayside ionosphere of
  {M}ars}.{\BBCQ}
\newblock
\APACjournalVolNumPages{Geophys. Res. Lett.}{39}{}{18202}.
\newblock
\begin{APACrefDOI} \doi{10.1029/2012GL053193} \end{APACrefDOI}
\PrintBackRefs{\CurrentBib}

\bibitem [\protect \citeauthoryear {%
Withers%
\ \protect \BOthers {.}}{%
Withers%
\ \protect \BOthers {.}}{%
{\protect \APACyear {2019}}%
}]{%
withers2019}
\APACinsertmetastar {%
withers2019}%
\begin{APACrefauthors}%
Withers, P.%
, Flynn, C\BPBI L.%
, Vogt, M\BPBI F.%
, Mayyasi, M.%
, Mahaffy, P.%
, Benna, M.%
\BDBL {}England, S.%
\end{APACrefauthors}%
\unskip\
\newblock
\APACrefYearMonthDay{2019}{}{}.
\newblock
{\BBOQ}\APACrefatitle {Mars's dayside upper ionospheric composition is affected
  by magnetic field conditions} {Mars's dayside upper ionospheric composition
  is affected by magnetic field conditions}.{\BBCQ}
\newblock
\APACjournalVolNumPages{J. Geophys. Res., In Press}{0}{ja}{}.
\newblock
\begin{APACrefURL}
  \url{https://agupubs.onlinelibrary.wiley.com/doi/abs/10.1029/2018JA026266}
  \end{APACrefURL}
\newblock
\begin{APACrefDOI} \doi{10.1029/2018JA026266} \end{APACrefDOI}
\PrintBackRefs{\CurrentBib}

\bibitem [\protect \citeauthoryear {%
{Withers}%
\ \protect \BOthers {.}}{%
{Withers}%
\ \protect \BOthers {.}}{%
{\protect \APACyear {2016}}%
}]{%
withers2016}
\APACinsertmetastar {%
withers2016}%
\begin{APACrefauthors}%
{Withers}, P.%
, {Matta}, M.%
, {Lester}, M.%
, {Andrews}, D.%
, {Edberg}, N\BPBI J\BPBI T.%
, {Nilsson}, H.%
\BDBL {}{Witasse}, O.%
\end{APACrefauthors}%
\unskip\
\newblock
\APACrefYearMonthDay{2016}{}{}.
\newblock
{\BBOQ}\APACrefatitle {{The morphology of the topside ionosphere of {M}ars
  under different solar wind conditions: {R}esults of a multi-instrument
  observing campaign by {M}ars {E}xpress in 2010}} {{The morphology of the
  topside ionosphere of {M}ars under different solar wind conditions: {R}esults
  of a multi-instrument observing campaign by {M}ars {E}xpress in
  2010}}.{\BBCQ}
\newblock
\APACjournalVolNumPages{Planet. Space Sci.}{120}{}{24-34}.
\newblock
\begin{APACrefDOI} \doi{10.1016/j.pss.2015.10.013} \end{APACrefDOI}
\PrintBackRefs{\CurrentBib}

\bibitem [\protect \citeauthoryear {%
Wu%
\ \protect \BOthers {.}}{%
Wu%
\ \protect \BOthers {.}}{%
{\protect \APACyear {2019}}%
}]{%
wu2019}
\APACinsertmetastar {%
wu2019}%
\begin{APACrefauthors}%
Wu, X\BHBI S.%
, Cui, J.%
, Xu, S\BPBI S.%
, Lillis, R\BPBI J.%
, Yelle, R\BPBI V.%
, Edberg, N\BPBI J\BPBI T.%
\BDBL {}Mitchell, D\BPBI L.%
\end{APACrefauthors}%
\unskip\
\newblock
\APACrefYearMonthDay{2019}{}{}.
\newblock
{\BBOQ}\APACrefatitle {The Morphology of the Topside Martian Ionosphere:
  {I}mplications on Bulk Ion Flow} {The morphology of the topside martian
  ionosphere: {I}mplications on bulk ion flow}.{\BBCQ}
\newblock
\APACjournalVolNumPages{J. Geophys. Res.}{124}{3}{734-751}.
\newblock
\begin{APACrefURL}
  \url{https://agupubs.onlinelibrary.wiley.com/doi/abs/10.1029/2018JE005895}
  \end{APACrefURL}
\newblock
\begin{APACrefDOI} \doi{10.1029/2018JE005895} \end{APACrefDOI}
\PrintBackRefs{\CurrentBib}

\bibitem [\protect \citeauthoryear {%
{Xu}%
\ \protect \BOthers {.}}{%
{Xu}%
\ \protect \BOthers {.}}{%
{\protect \APACyear {2019}}%
}]{%
xu2019}
\APACinsertmetastar {%
xu2019}%
\begin{APACrefauthors}%
{Xu}, S.%
, {Curry}, S\BPBI M.%
, {Mitchell}, D\BPBI L.%
, {Luhmann}, J\BPBI G.%
, {Lillis}, R\BPBI J.%
\BCBL {}\ \BBA {} {Dong}, C.%
\end{APACrefauthors}%
\unskip\
\newblock
\APACrefYearMonthDay{2019}{}{}.
\newblock
{\BBOQ}\APACrefatitle {{Magnetic Topology Response to the 2003 {H}alloween
  {ICME} Event at Mars}} {{Magnetic Topology Response to the 2003 {H}alloween
  {ICME} Event at Mars}}.{\BBCQ}
\newblock
\APACjournalVolNumPages{J. Geophys. Res.}{124}{1}{151-165}.
\newblock
\begin{APACrefDOI} \doi{10.1029/2018JA026118} \end{APACrefDOI}
\PrintBackRefs{\CurrentBib}

\bibitem [\protect \citeauthoryear {%
{Xu}%
\ \protect \BOthers {.}}{%
{Xu}%
\ \protect \BOthers {.}}{%
{\protect \APACyear {2018}}%
}]{%
xu2018}
\APACinsertmetastar {%
xu2018}%
\begin{APACrefauthors}%
{Xu}, S.%
, {Fang}, X.%
, {Mitchell}, D\BPBI L.%
, {Ma}, Y.%
, {Luhmann}, J\BPBI G.%
, {DiBraccio}, G\BPBI A.%
\BDBL {}{Lee}, C\BPBI O.%
\end{APACrefauthors}%
\unskip\
\newblock
\APACrefYearMonthDay{2018}{}{}.
\newblock
{\BBOQ}\APACrefatitle {{Investigation of Martian Magnetic Topology Response to
  2017 {S}eptember {ICME}}} {{Investigation of Martian Magnetic Topology
  Response to 2017 {S}eptember {ICME}}}.{\BBCQ}
\newblock
\APACjournalVolNumPages{Geophys. Res. Lett.}{45}{15}{7337-7346}.
\newblock
\begin{APACrefDOI} \doi{10.1029/2018GL077708} \end{APACrefDOI}
\PrintBackRefs{\CurrentBib}

\end{thebibliography}




    \newpage

    \begin{figure}[h]
    \centering
    \includegraphics[width=0.5\textwidth]{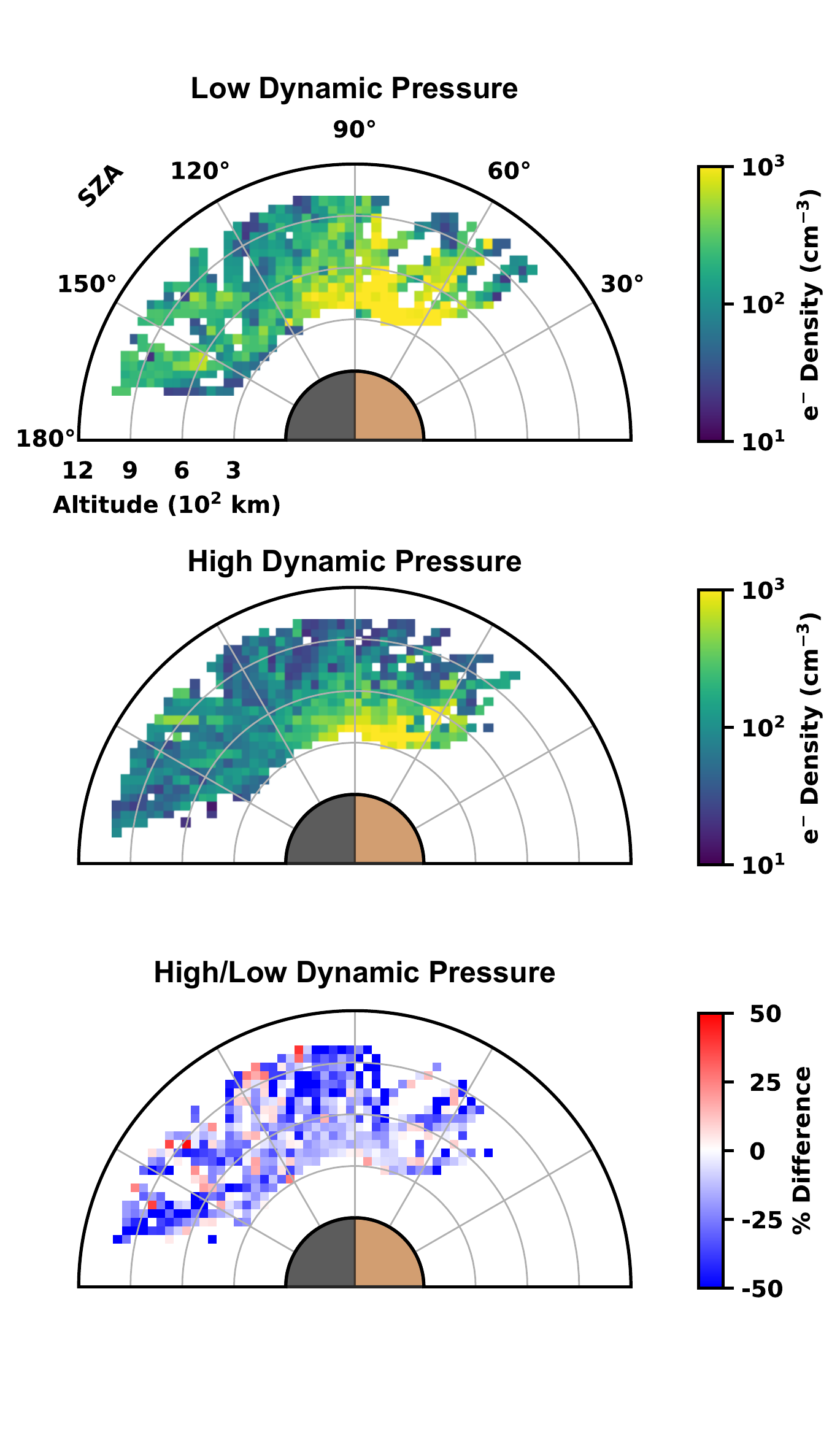}
    \caption{Electron densities in the topside ionosphere are depleted at all solar zenith angles (SZA) during times of high solar wind dynamic pressure. The top two panels show electron densities as a function of SZA and altitude for low (top) and high (middle) solar wind dynamic pressures. The bottom panel shows the percent difference between the two panels with negative values (blue colors) indicating that electron densities are lower during high dynamic pressure. Note that Mars is not drawn to scale in order to emphasize ionospheric patterns. }
    \label{pressure}
    \end{figure}
    
\end{document}